\documentclass[a4paper,11pt]{article}
\usepackage{jinstpub} 
\usepackage{lineno}

\title{\boldmath Study of silicon photomultipliers for the readout of a lead/scintillating-fiber calorimeter}

\author[a,b]{F.~Alemanno,}
\author[a,b]{P.~Bernardini,}
\author[b]{A.~Corvaglia,}
\author[a,b]{G.~De~Matteis,}
\author[a,b]{L.~Martina,}
\author[b]{A.~Miccoli,}
\author[a,b]{M.~Panareo,}
\author[b]{M.P.~Panetta,}
\author[a,b]{C.~Pinto,}
\author[b,1]{A.~Surdo\note{Corresponding author.}}

\affiliation[a]{Dipartimento di Matematica e Fisica, Universit\`a del Salento, Lecce, Italy }
\affiliation[b]{Istituto Nazionale di Fisica Nucleare, Lecce, Italy}

\emailAdd{antonio.surdo@le.infn.it}

\abstract{The KLOE electromagnetic calorimeter is expected to be reused in the Near Detector complex of the DUNE experiment at Fermilab. The possible substitution of traditional Photomultiplier Tubes (PMTs) with Silicon Photomultipliers (SiPMs) in the refurbished calorimeter is the object of this investigation. A block of the KLOE lead-scintillating fiber calorimeter has been equipped with light guides and external trigger scintillators. The signals induced by cosmic rays and environmental radioactivity have been collected by SiPM arrays on one side of the calorimeter, and by conventional PMTs on the opposite side. Efficiency, stability, and timing resolution of SiPMs have been studied and compared with KLOE-PMTs performance. Conclusions about the convenience of substituting PMTs with SiPMs are drawn. }


\begin{document}
\maketitle
\flushbottom

\section{Introduction}
\label{sec:intro}

Silicon Photomultipliers (SiPMs) are solid-state photodetectors~\cite{bib:acerbi}, which consist of a high-density matrix of SPADs (Single-Photon Avalanche Diodes) operating in Geiger mode as independent detectors. The number of fired photodiodes is proportional to the number of collected photons.
The SiPM performance improved continuously in the last years and presently SiPMs are widely used in physics instrumentation, ranging from accelerator to astroparticle physics experiments. In particular, in spite of some drawbacks (small active area, limited dynamic range, strong dependence on temperature), they have specific advantages for calorimetry applications~\cite{bib:biol}. First of all, the SiPM photodetection range is compatible with the typical wavelength-shifted light of the scintillating fibers ($\sim$~460~nm), and unlike the PMTs, SiPMs are insensitive to magnetic fields. Furthermore, since SiPMs operate at low voltage, the high voltage power supply would not be necessary, with convenience in compactness and cost.

The present study aims to evaluate the compatibility of SiPM readout of the KLOE electromagnetic calorimeter~\cite{bib:kloe} and the possible capability of SiPMs to allow improvements in efficiency and timing resolution over standard PMT readout. This check is requested because the KLOE calorimeter is going to be refurbished as an element of SAND (System for on-Axis Neutrino Detection) in the Near Detector complex~\cite{bib:ND-CDR} of the DUNE experiment~\cite{bib:dune}.

\section{Experimental setup}\label{sec:exp}
\begin{figure}[htbp]
\centering
\includegraphics[width=0.7\textwidth]{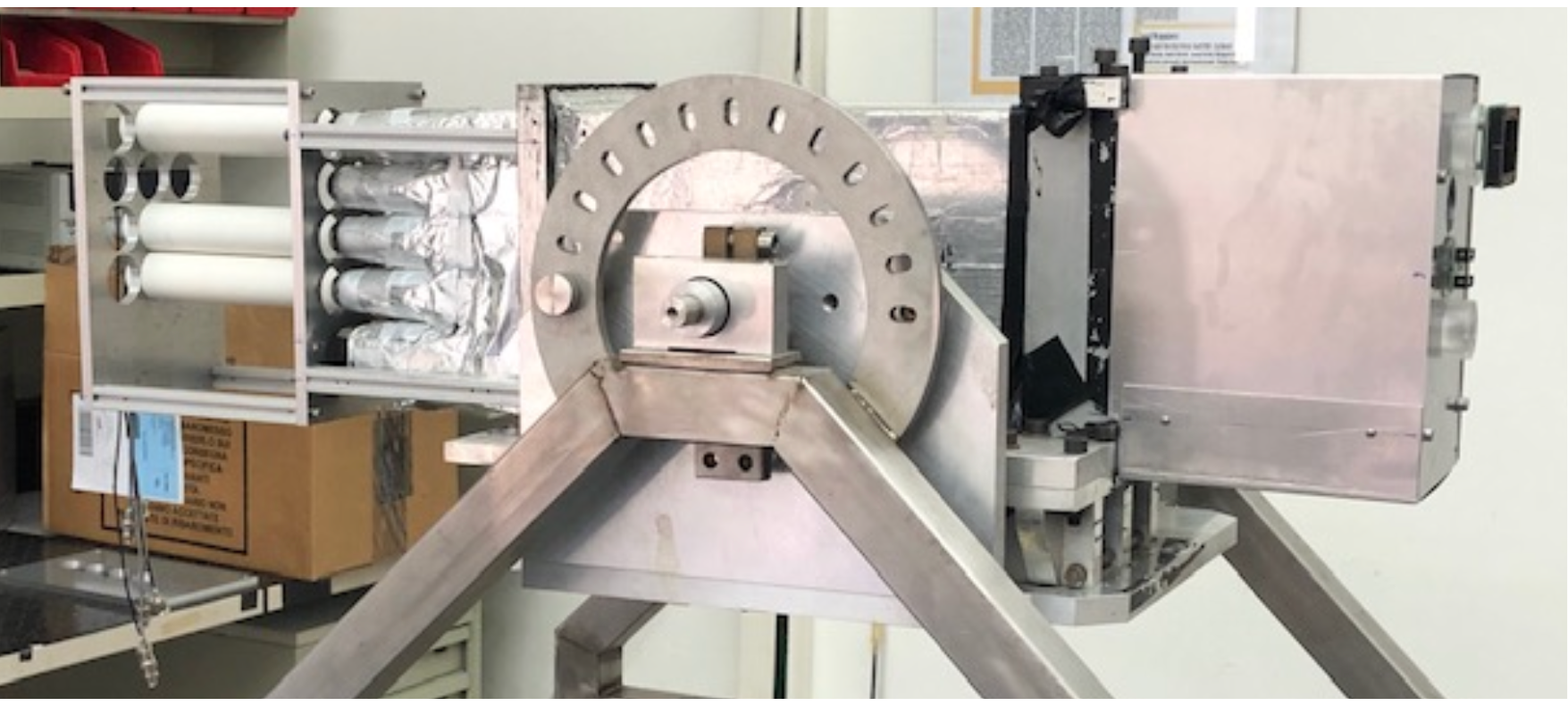}
\vspace{-0.2cm}
\caption{Picture of the experimental setup. The ECAL is in the center. The PMTs with the light guides are visible on the left side while the SiPMs are attached to the box containing the light guides on the right side.}
\label{fig:picture}
\end{figure}

\begin{figure}[htbp]
\centering
\includegraphics[width=0.9\textwidth]{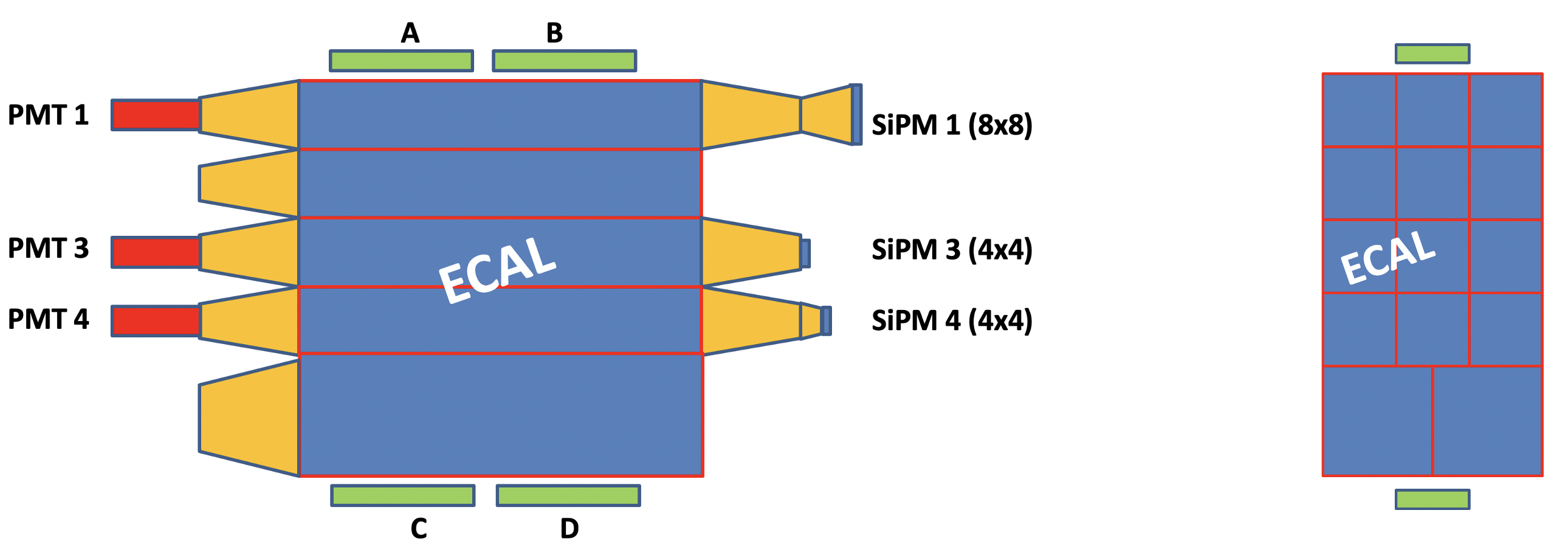}
\caption{Representation of the experimental setup front (left) and side (right) view, not to scale. The lead-fibers stack (ECAL) is divided into 14 cells represented by the central blue elements. The instrumented photosensors are shown on both sides of each cell of the central column and indicated in red for the PMTs and blue for the SiPM arrays, connected with the ECAL through light guides (in yellow). The scintillator counters (A, B, C, D) are placed on the top and bottom of the setup to provide the external trigger as illustrated by green rectangles.}
\label{fig:set}
\end{figure}

The experimental setup used in this study is shown in Fig.~\ref{fig:picture} (photo) and  Fig.~\ref{fig:set} (schematic view). 


The principal element (ECAL) is a cut-out of the KLOE calorimeter~\cite{bib:kloe}, already used in another test~\cite{bib:mazzitelli}. It is composed of a stack of thick grooved lead foils, alternating with layers of scintillating fibers (blue-green, type Pol.Hi.Tech-0046) with the peak emission wavelength $\lambda_{\mathrm{peak}} \approx 460$~nm. The composite density results in 5 g/cm$^3$.

The upper part of the ECAL volume is segmented into 12 cells (section $4.4 \times 4.4$ cm$^2$, horizontal length 40.0 cm), while two larger cells (section $6.6 \times 6.6$ cm$^2$) occupy the lower part. This results in a total ECAL thickness of $\approx 15~X_0$ ($\approx 2.7~X_0$ for each smaller cell), considering a particle vertically crossing the calorimeter from top to bottom. 

\begin{figure}[htbp]
\centering 
\includegraphics[width=0.412\textwidth]{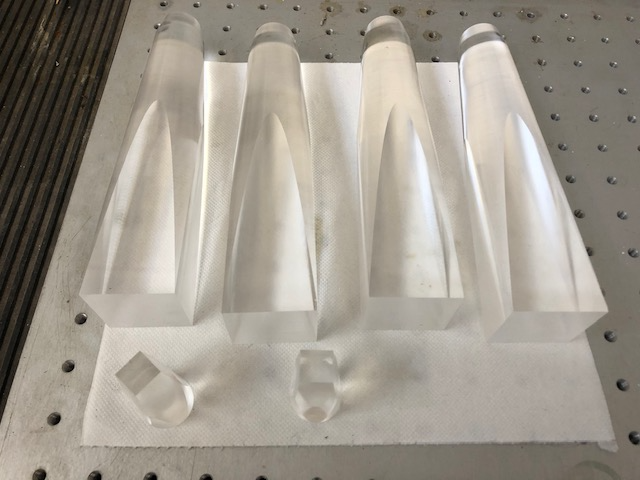}
\includegraphics[width=0.45\textwidth]{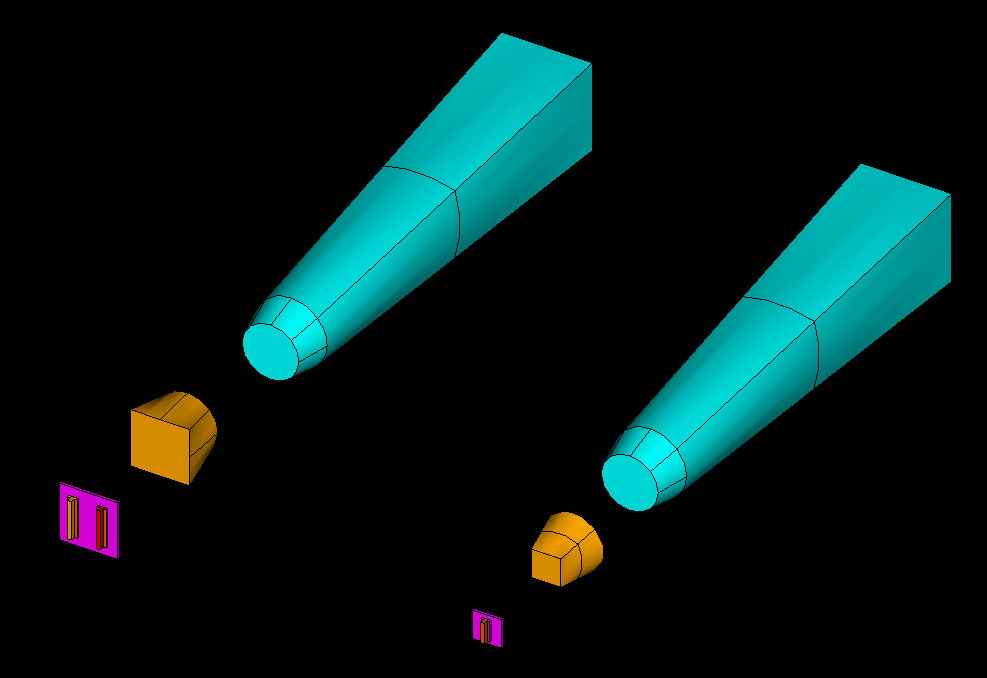}
\caption{Lucite light guides used to couple PMTs and SiPMs with the calorimeter cells connecting 
the square section of the calorimeter bars with the circular one of the photodetectors. Left: The long light guides are used for all the PMTs and SiPMs, while the short ones act as adapters for SiPM 1 and SiPM 4, in order to couple different surfaces and allow the maximum light collection. Right: design of light guides (azure), adapters (orange), and SiPM arrays (magenta).} 
\label{fig:LightGuides} 
\end{figure}

Lucite light guides (Fig.~\ref{fig:LightGuides}), similar to the ones used in KLOE, are glued at both edges of the cells in the central column. In order to obtain a good light collection while optimizing the coupling between ECAL and photodetectors, the light guides combine a tapered mixing part with a quadrangular entrance and circular exit terminating with a Winston cone (exit section $\approx500$ mm$^2$). On one side (left in Fig.~\ref{fig:set}) they are coupled to Hamamatsu-R5946 1.5' photomultipliers~\cite{bib:hama_PMT}, with a cathode properly fitting the circular base. On the opposite side, the readout is performed using arrays of $8\times8$ or $4\times4$ SiPMs (Hamamatsu S13361-3050 series~\cite{bib:hama_SiPM}). Since the substitution of the 4880 KLOE light guides is an inconvenient and complex operation, the SiPMs were chosen in order to cover as much as possible of the present KLOE light guide surface. It was evaluated that a custom SiPM shaped to fit the circular guides is not strictly necessary for these measurements. Consequently, additional coupling devices are adopted to associate the SiPM arrays with the circular exit of the light guides to account for the different SiPM surfaces and to test possible minor adjustments of the KLOE-like guides. The coupling solutions are shown in Figs.~\ref{fig:set} and \ref{fig:LightGuides}, and described here:
\begin{itemize}
\item[-] SiPM 1, a lucite adapter is glued to fit the $8\times8$ SiPM array (area 666~mm$^2$),
\item[-] SiPM 2, not implemented,
\item[-] SiPM 3, the $4\times4$ array is directly coupled to the light guide (covered fraction $\sim$170/500),
\item[-] SiPM 4, a lucite adapter is glued to fit the $4\times4$ SiPM array (area 170~mm$^2$).
\end{itemize}

\noindent The light guides and the adapters are coupled with the calorimeter using the Polytec EP 601-LV optical cement (refractive index 1.53), and with PMTs and SiPMs through a silicon paste (optical coupling compound `Dow Corning 20-057'). Moreover, a system of adjustable springs pushes gently the PMTs (according to the KLOE design) and the SiPM arrays toward the light guides. 
A CAEN FERS-DT5202 electronic board~\cite{bib:caen} with its readout software is used to characterize the SiPM performance (Sec.~\ref{sec:chara}) and to select events for the efficiency measurement (Sec.~\ref{sec:eff}). The latter measurement is based on an external trigger for cosmic rays made by 4 plastic scintillator bars read out by PMTs (green elements in Fig.~\ref{fig:set}), placed on top (A, B) and bottom (C, D) of the experimental setup. A Teledyne Lecroy Waverunner 640Zi oscilloscope is employed for both SiPM and PMT data taking, along with CAEN and Lecroy modules, to measure the timing resolutions (Sec.~\ref{sec:time}).

\subsection{SiPMs and PMTs}
The SiPMs used in this test are the MPPC arrays of the Hamamatsu S13361-3050 series. Both the available configurations ($4\times4$ and $8\times8$ channels) were mounted but only the performance of 16-channels arrays (SiPM 3 and 4) are reported in this paper~\footnote{ The $8\times8$ array and its readout printed circuit are bulky and difficult to fit in the KLOE light guides assembling setup.}. This specific MPPC series is chosen in order to achieve the maximum Photo-Detection Efficiency (PDE$_{MAX}$) close to the peak wavelength of the scintillating fibers (typically PDE$_{MAX} = 40\%$ at $\lambda = 450$ nm, ranging from $\sim$30\% to $\sim$60\% as a function of the overvoltage). Furthermore, the SiPM arrays are characterized by a large effective photosensitive area of $3\times3$ mm$^2$ for each channel (fill factor $= 74\%$ ), a 50 $\mu$m pixel size which ensures a wide dynamic range of photon counting ($3,584$ pixels for each channel), and a gain of $O(10^6)$. Other features are summarized in~\cite{bib:hama_SiPM}. The SiPMs readout is performed with a desktop front-end unit, CAEN FERS-DT5202, housing Weeroc Citiroc-1A chips~\cite{bib:c_rock}, which hosts a SiPM power supply.
As indicated before, in this test the SiPM arrays were chosen in order to maximize the coverage of the light guide. Since it's excluded to substitute the single PMT KLOE channel with a larger number of readout channels, the SiPM array is considered as one element in the following measurements. Moreover, this arrangement makes the measurements analogous to that of the PMT. The SiPM performance are compared with that of the Hamamatsu R5946 1.5’ mesh photomultiplier already used in the KLOE calorimeter. The quantum efficiency of this PMT is 18\% at $\lambda = 450$~nm, according to the device datasheet. 

\section{SiPM characterization and light yield}\label{sec:chara}
\begin{figure}[ht!]
\centering 
\includegraphics[width=0.45\textwidth]{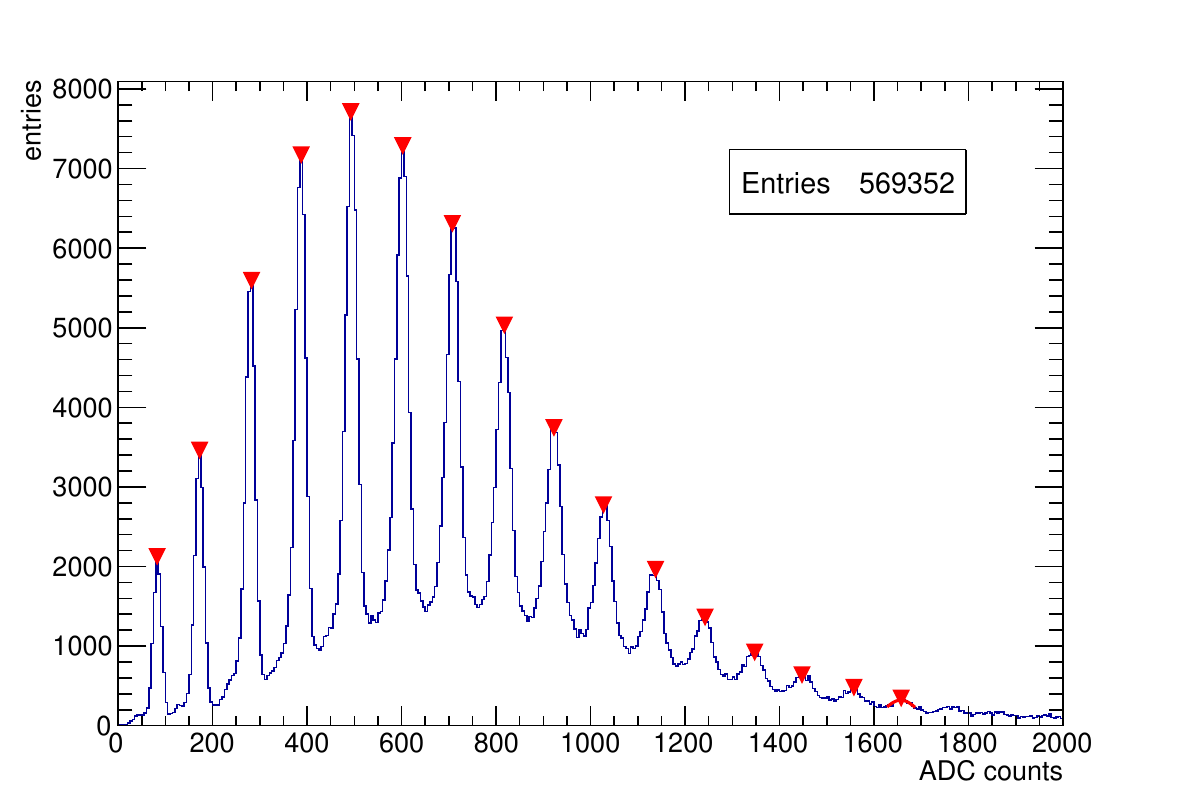}
\includegraphics[width=0.45\textwidth]{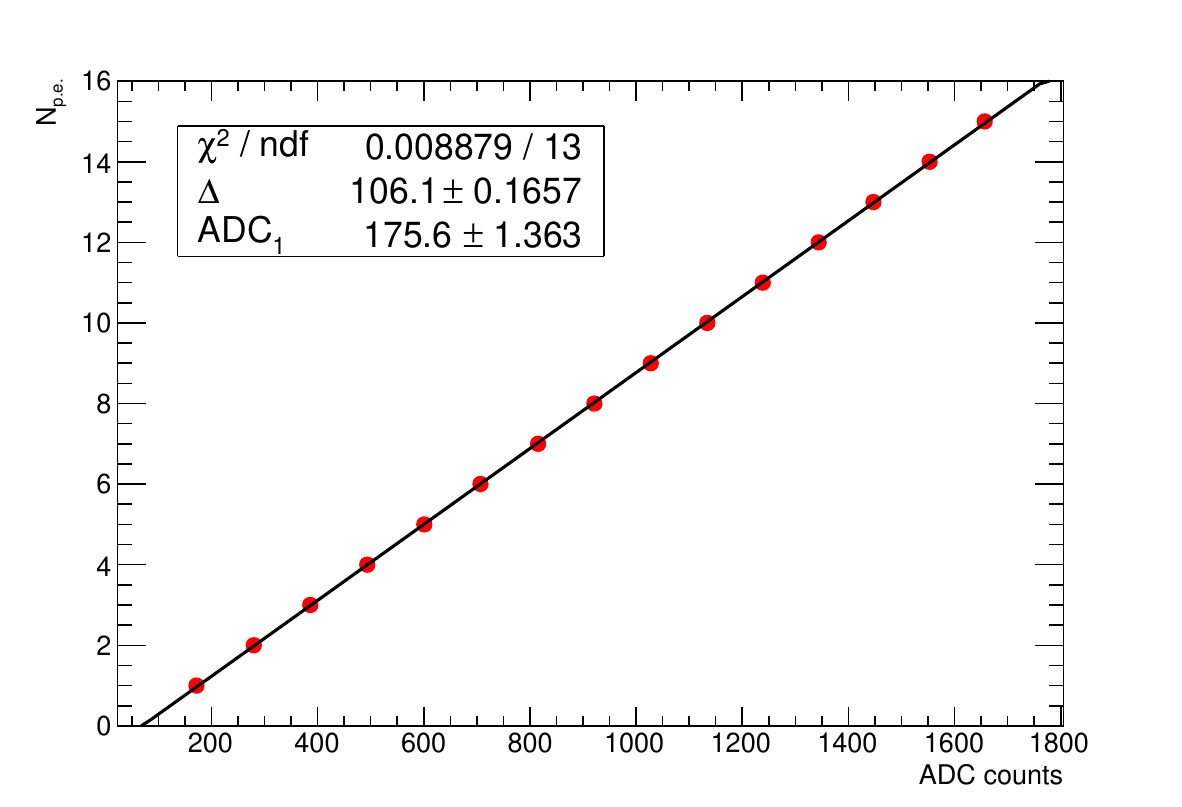}
\caption{Left: photoelectron spectrum of SiPM 3 (entries from each single channel). The spectrum is truncated at 2000 counts and the first peak is the pedestal. The peaks, recognized by the Root TSpectrum method~\cite{bib:root}, are marked by red triangles. Right: number of collected photoelectrons versus the peak ADC counts. The pedestal is excluded and a linear fit is superimposed according to formula~(\ref{eq:ngamma}).} 
\label{fig:spectrum106} 
\end{figure}

The known dependence of the SiPM performance on the temperature~\cite{bib:temperatura,bib:web} required to work at stable conditions ($\sim23^\circ$C). The SiPM calibration is made in a self-trigger acquisition configuration requiring at least one channel over the threshold. Initially, the SiPM channels are calibrated individually by voltage (V$_{BIAS}$) regulation, in order to get the same fixed rate ($\approx 3.5$ kHz) for each element of the array. In this case, the signal in the SiPMs is generated by dark counts, cosmic charged particles and environmental radioactivity. 
Afterwards, in order to select mainly charged particles crossing the calorimeter, the data are taken requiring the self-trigger on the 16 channels of each SiPM along with the AND of PMT~3 and PMT~4 (trigger rate $\approx$ 1.4~Hz). The result is shown in Fig.~\ref{fig:spectrum106} left, where the histogram is filled with the ADC counts taken from each single channel of the SiPM~3. The same procedure is applied to obtain a similar spectrum for the SiPM~4. 

The calibration works properly for all channels of both SiPMs. Indeed, the peaks are placed in the same position (same ADC value) for all of them and we do not observe any broadening of the peaks when many channels are plotted together. Moreover, the SiPM channels show linearity and stability up to a 15-photoelectrons (p.e.) signal, as can be seen in Fig.~\ref{fig:spectrum106} right. From the ADC values of the peaks, it is possible to infer the formula to convert the ADC signal in number of p.e. expressed as follows:
\begin{equation}
N_{p.e.} = 1\ +\ \frac{ADC - ADC_1}{\Delta},
\label{eq:ngamma}
\end{equation}
\noindent where $ADC_1$ is the ADC value for one photon (first peak after the pedestal) and $\Delta$ is the ADC-difference between adjacent peaks. This formula has been applied successfully to fit the number of p.e. associated with each peak and the ADC counts of the peak (Fig.~\ref{fig:spectrum106}, right). Saturation effects are not visible up to 15 p.e.

\begin{figure}[ht!]
\centering
\includegraphics[width=0.45\textwidth]{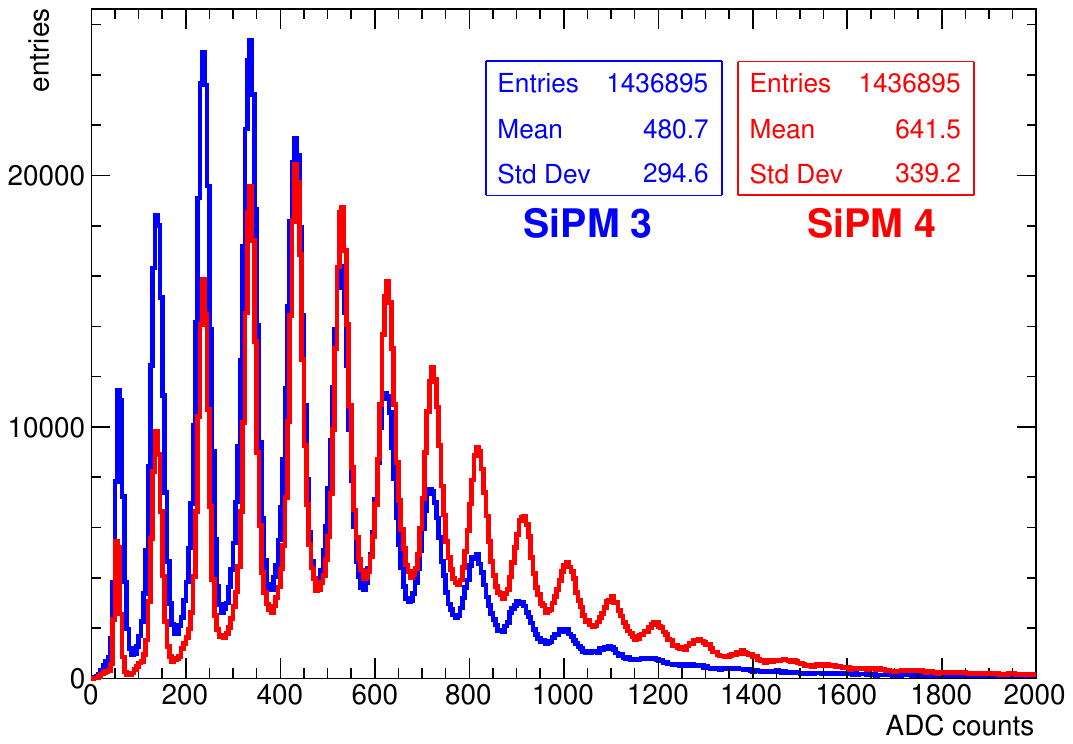}
\includegraphics[width=0.50\textwidth]{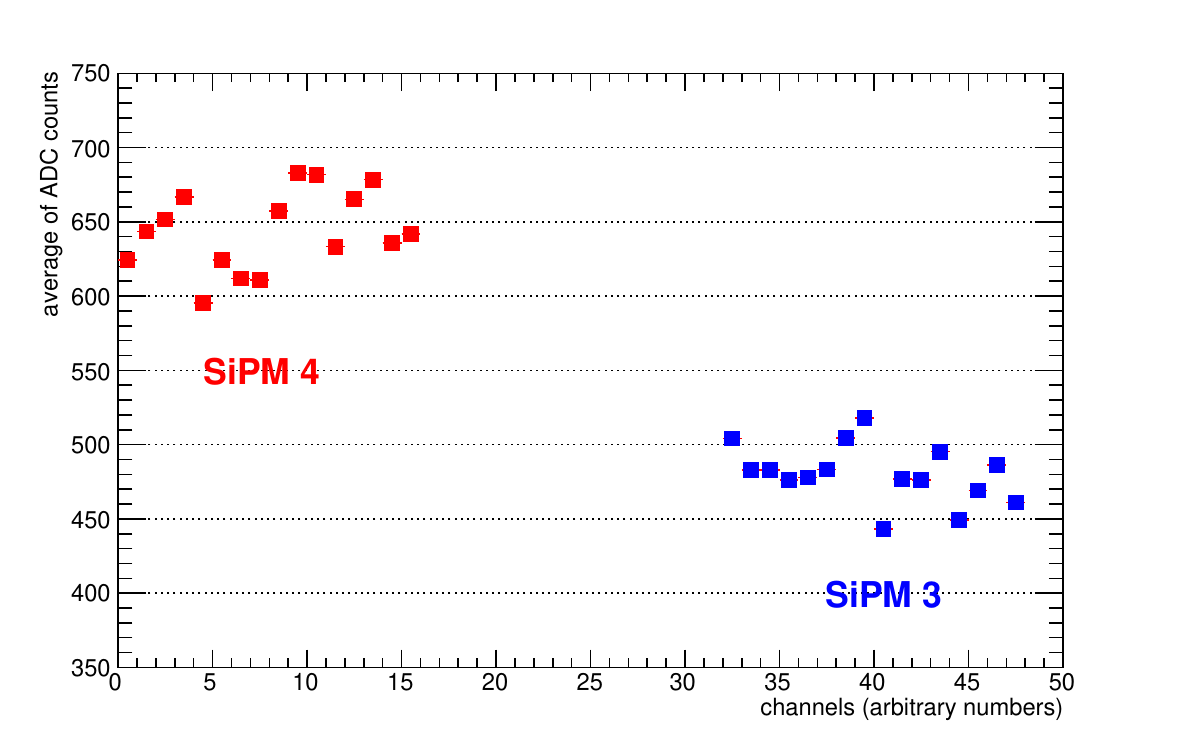}
\caption{Left: the ADC single-channel spectrum for SiPM 3 (blue), compared with that of SiPM 4 (red).
The spectra are shown up to 2000 counts. 
Right: average ADC counts versus arbitrary channel numbers. The average values are calculated from the whole distribution.} \label{fig:uniform} 
\end{figure}

From the comparison of the ADC single-channel spectra of SiPM 3 and 4, it is possible to check the performance of the adapter glued to the Winston cone. For this check, the events were acquired when at least one channel was over the threshold in both SiPMs. In such trigger configuration, the acquired events are mainly due to cosmic particles entering both calorimeter modules, resulting in a rate of $\approx$ 1.5~Hz. In Fig.~\ref{fig:uniform} left, the two spectra are displayed. It is evident that the light collection is higher for the coupling with the adapter (SiPM 4). Indeed, using Equation~(\ref{eq:ngamma}) and the parameters of the fit in Fig.~\ref{fig:spectrum106}, right, the mean number of collected p.e. in these 
distributions is $\sim3.9$ for SiPM 3 and $\sim5.4$ for SiPM 4. This is the expected result considering the optical coupling improvement due to the adapter used for SiPM 4 (the adapter surface perfectly matches the active area of the SiPM).

This result is also confirmed by the observation that the average ADC counts for all the channels of SiPM 4 are higher with respect to those of SiPM 3 (Fig.~\ref{fig:uniform}, right). The light collection is quite uniform and a range of $\sim$100 is the spread of the average ADC counts for both SiPM 3 and SiPM 4. Using the fit in Fig.~\ref{fig:spectrum106}, right, it is estimated that this spread corresponds to $\pm$0.5 p.e. Therefore all channels have quite similar efficiency in the same SiPM and the Winston cones homogenize the light signal. Using light guides on both ends of each calorimeter module ensures that the SiPM configuration is comparable to the PMT one in terms of light collection. Moreover, it was verified that both SiPM 3 and SiPM 4 are well-suited for the intended study, and they perform similarly.

\section{Efficiency}\label{sec:eff}

The SiPM efficiency is measured by detecting cosmic muons at a stable lab temperature of 23$\pm1^\circ$C. A system of 4 scintillators, with an area of $\sim 2 \times 7.5$ cm$^2$ each, provides the external trigger. Two scintillators are placed above (A and B green elements in Fig.~\ref{fig:set}) and two below (C and D in Fig.~\ref{fig:set}) the calorimeter, overlapping with its central modules. The trigger logic consists of a fired scintillator on the top (A or B) and a fired scintillator on the bottom (C or D) in a time window of 30~ns. The measured rate is 2.6 mHz as expected from the cosmic muon flux, taking into account the narrow solid angle due to the trigger geometry. When a cosmic muon activates this trigger, it is expected to pass through the stack of ECAL modules, generating photons in the fibers that can be detected by SiPMs and PMTs. Thus, during the SiPMs data acquisition, the PMT efficiency has also been measured. Finally, we observe that the particle-crossing probability is symmetrical with respect to the center of ECAL modules and the trigger does not favor SiPMs over PMTs or vice versa.

The PMT signals are counted by a CAEN Quad Scaler module after they are shaped with a threshold of 40 mV. The SiPM signals are managed by the CAEN FERS-DT5202 board with two different outputs, both requiring the self-trigger and validated by the external trigger. The used self-trigger is the OR-16 trigger with a low threshold of 210 a.u., to collect also events with only 1 photon on the whole SiPM array. 
When the validation from the external trigger is satisfied, the counts are acquired with the CAEN Quad Scaler module and the digital full information on the SiPM signals is collected by the data acquisition program running on a desktop personal computer. Although the requirements are identical, the counts from the outputs of the FERS board are different. The events collected by the acquisition program (N$_{DAQ}$) are typically fewer than the shaped signals (k$_{SiPM}$) sent directly to the Quad Scaler as an effect of the DAQ board dead time. Indeed the board manages all OR-16 triggers,
so an event associated with the external trigger can be lost if it arrives when the board is busy with the analog-to-digital conversion of a previous event. On the contrary, this dead time does not affect the direct output and the consequent counting by the scaler.

\begin{figure}[htbp]
\centering
\includegraphics[width=0.43\textwidth]{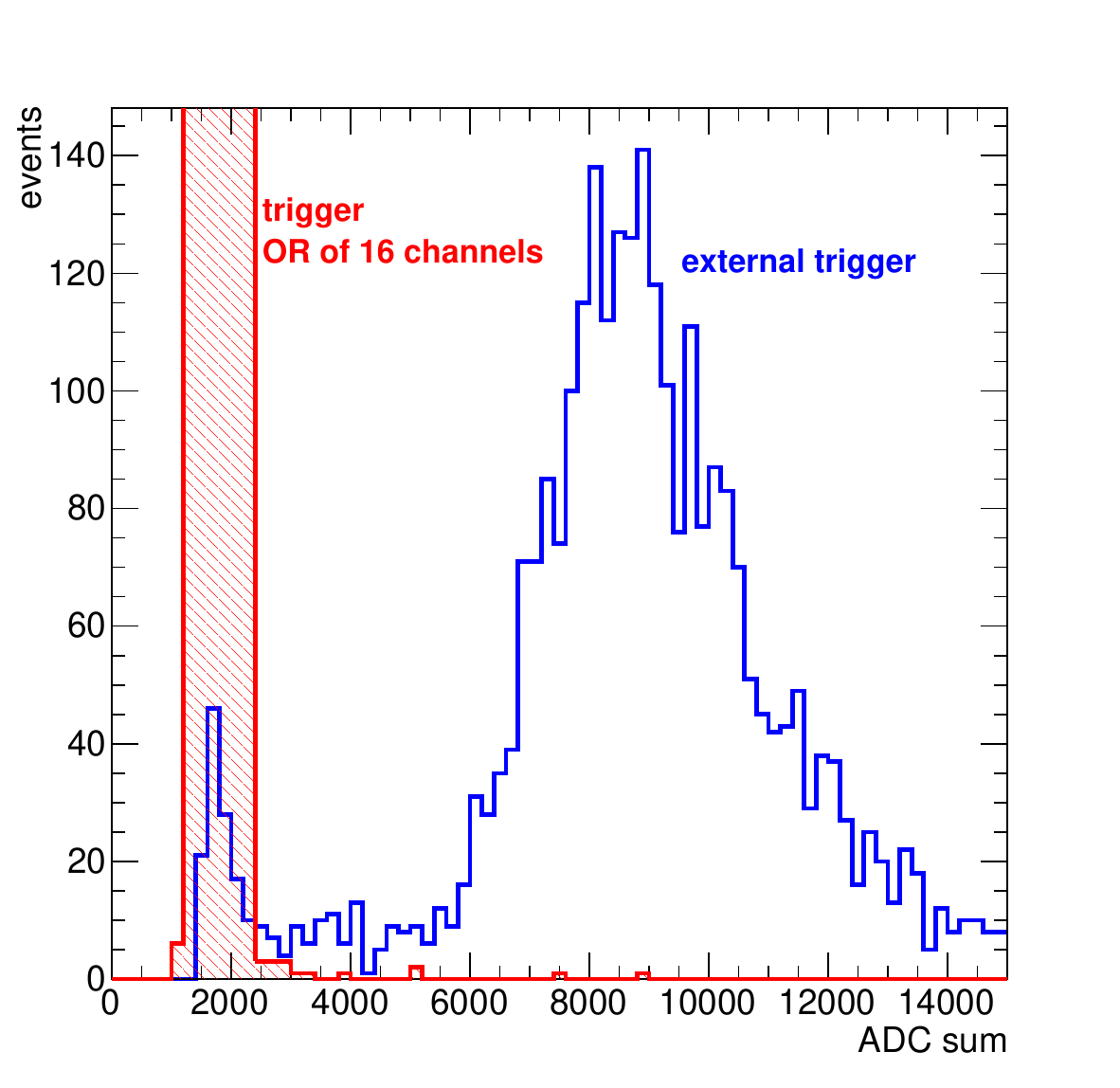}
\includegraphics[width=0.43\textwidth]{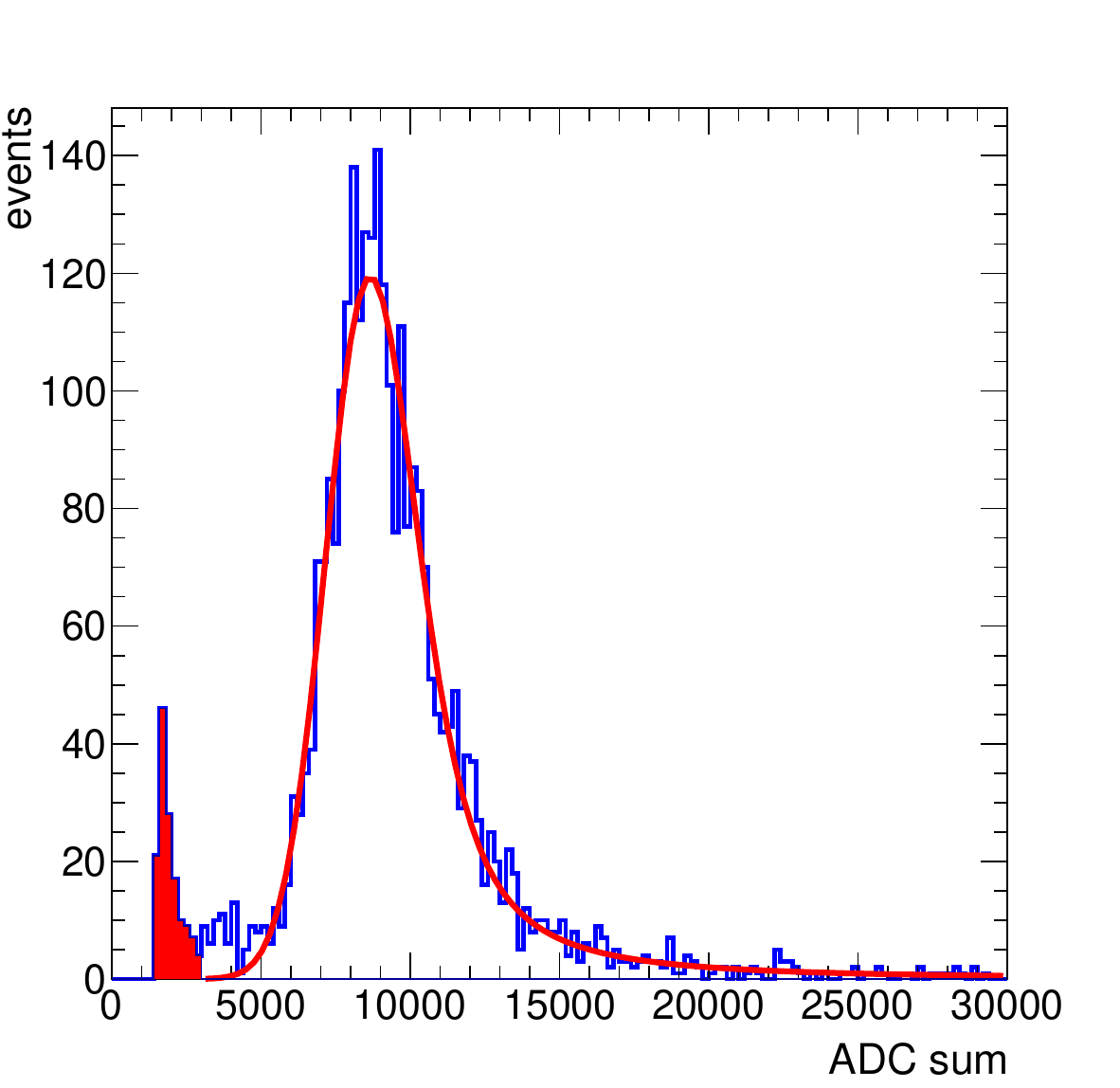}
\hspace{0.1cm}
\caption{Sum of the ADC values (ADC-sum) for SiPM 3. Left: Distributions for different trigger conditions in the range 0-15000 (more details in the text). Right: Distribution with external trigger in the range 0-30000. The superimposed fit is a combination of Landau and Gaussian functions (LanGaus). The red part of the distribution is rejected requiring ADC-sum $>$ 3000.} 
\label{fig:sum} 
\end{figure}

The event sample collected by the DAQ is analyzed in order to identify and remove a possible residual background due to the dark count induced by the choice of a very low threshold. Before the definition of a cut to clean the sample, it must be taken into account that the SiPM array is studied as a unique detector, and then the cut must be applied to the sum of the ADC of all channels (ADC-sum). To investigate the SiPM noise, samples have been collected requiring only the OR-16 trigger and not the validation of the external trigger. The ADC-sum distribution of the sample collected in this condition for SiPM 3 is shown in Fig.~\ref{fig:sum}, left (red histogram). A clear peak is visible at $\sim$2000~ADC with some events at higher values. In the same plot also the distribution for events validated by the external trigger is reported. It is evident the superposition of the two distributions at low values of the ADC-sum. Similar results were obtained for the SiPM 4.

On this basis, a reasonable cut to measure the SiPM efficiency is ADC-sum $>$ 3000. Another more stringent cut is suggested by the fit of the ADC-sum distribution by means of a Landau function convoluted with a Gaussian (Fig.~\ref{fig:sum}, right) to take into account the multiplicity spread of collected photons and the behavior of the detector. In this case, the samples can be reduced by requiring ADC-sum $>$ 5000. Then, the SiPM efficiency calculated from the scaler counts ($\epsilon$) must be reduced by accepting only the fraction of events surviving the ADC-sum cut ($\epsilon'$). The corrected SiPM efficiency can be expressed as:

\begin{equation}
E_{SiPM} = \epsilon \ \epsilon' = \frac{k_{SiPM}}{N_{SiPM}} \ \frac{k_{DAQ}}{N_{DAQ}},
\label{eq:esipm}
\end{equation}
\noindent where $k_{SiPM}$ is the number of SiPM events on the scaler, $N_{SiPM}$ is the number of external triggers, 
$k_{DAQ}$ and $N_{DAQ}$ are respectively the number of selected events and the total number of events in the sample 
collected by the DAQ. On the other side, the PMT efficiency is calculated simply from the scaler counts:

\begin{equation}
E_{PMT} = \frac{k_{PMT}}{N_{PMT}},
\label{eq:epmt}
\end{equation}
\noindent where the meaning of $k_{PMT}$ and $N_{PMT}$ is obvious. The results of the efficiency measurement are shown in Table~\ref{tab:eff12} where the Clopper-Pearson method~\cite{bib:cp} has been used to estimate the errors and 3000 is selected as the ADC-sum threshold to accept SiPM events. It can be observed that the higher average signal for SiPM 4 (see again Fig.~\ref{fig:uniform}) 
is related to a higher value of $\epsilon$ (96.7\%) with respect to SiPM 3 (95.0\%). This difference is balanced by a higher contribution of noise events and a lower value of $\epsilon'$ (93.9\%) compared to SiPM 3 (95.5\%). The resulting efficiencies ($\epsilon\epsilon'$) for SiPM 3 and 4 are finally very close as a demonstration of the method reliability. As an additional test, a different self-trigger condition was set in the CAEN FERS-DT5202 board by increasing the threshold to 230. The result was a reduction of $\epsilon$ and an enhancement of $\epsilon'$ without any significant change in the final result.

Using the stricter threshold suggested by the LanGaus fit (ADC-sum $>$ 5000), the efficiencies are 88.34\% and 89.16\% for SiPM 3 and 4, respectively, with errors of the same magnitude as in Table~\ref{tab:eff12}.  In conclusion, the efficiencies of PMTs and SiPMs are very close, with the PMT efficiencies being slightly higher than the SiPM ones, essentially as a result of a higher noise level of the latter.

\begin{table}[htbp]
\centering
\caption{Efficiency measurements for PMTs and SiPMs. The k$_{DAQ}$ value is estimated requiring ADC-sum $>$ 3000.}\label{tab:eff12} 
\smallskip
\begin{tabular}{c c c c c c} \hline
\textit{PMT}    & k$_{PMT}$ & N$_{PMT}$ &          &          & E$_{PMT}$ (\%)          \\ \hline
                &           &           &          &          &                         \\
\textit{3}      &   7010    &  7615     &          &          & $92.06^{+0.14}_{-0.15}$ \\
                &           &           &          &          &                         \\
\textit{4}      &   6739    &  7392     &          &          & $91.17^{+0.15}_{-0.16}$ \\
                &           &           &          &          &                         \\ \hline
\textit{SiPM}   &k$_{SiPM}$ & N$_{SiPM}$& k$_{DAQ}$& N$_{DAQ}$& E$_{SiPM}$ (\%)         \\ \hline
                &           &           &          &          &                         \\
\textit{3}      &  4324     &  4552     &  2999    &   3141   & $90.70^{+0.22}_{-0.23}$ \\
                &           &           &          &          &                         \\
\textit{4}      &  4820     &  4982     &  2680    &   2855   & $90.82^{+0.23}_{-0.25}$ \\   
                &           &           &          &          &                         \\ \hline
\end{tabular}
\end{table}


\section{Timing resolution}\label{sec:time}
\begin{figure}[ht!] \centering 
\includegraphics[width=0.5\textwidth]{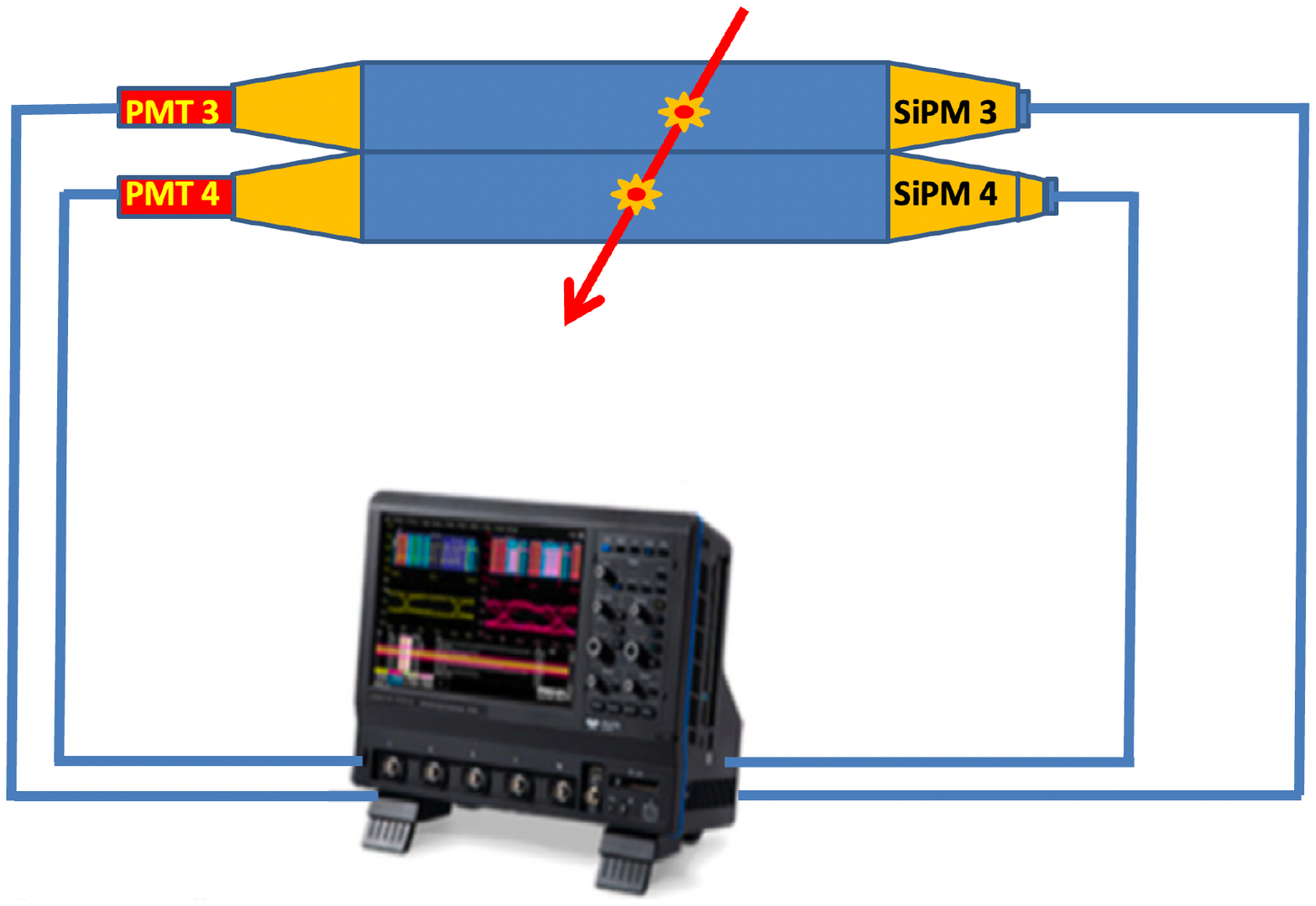}
\caption{Setup used to measure the timing resolution of SiPMs and PMTs. Two ECAL modules, coupled with PMTs on one side and SiPMs on the opposite side, are used for this measurement. The signal is acquired by the oscilloscope, which also provides the timing information.}
\label{fig:set_det} \end{figure}

The timing resolution of the photodetectors has been measured using the rising time at 50\% of the signal amplitude. The experimental setup is shown in Fig.~\ref{fig:set_det}, where the ECAL modules 3 and 4 are coupled with PMTs on one side and SiPMs on the opposite side, both connected to the oscilloscope. 

\begin{figure}[ht!] \centering
\includegraphics[width=0.99\textwidth]{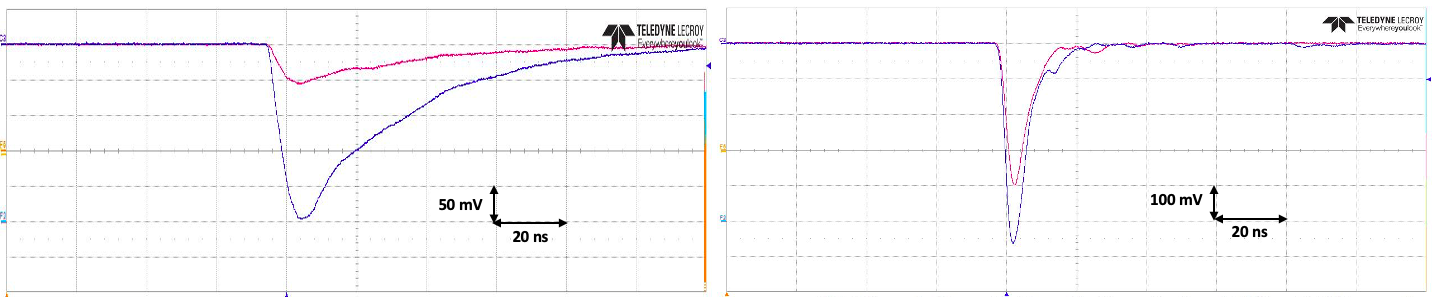}
\caption{Coincident signals on the oscilloscope display using two SiPMs (left) or two PMTs (right).
The trigger for the oscilloscope acquisition is generated by the request of a signal amplitude from the two channels over a properly set threshold, both for SiPM (30 mV threshold) and PMT case (100 mV threshold).}
\label{fig:oscillo} \end{figure}

In the case of PMTs, the signals from modules 3 and 4 are compared in order to measure $\Delta t$. For SiPMs, the measurement is carried out considering either two channels from the same array (SiPM~3~$\alpha$ and SiPM~3~$\beta$ or SiPM~4~$\alpha$ and SiPM~4~$\beta$, with $\alpha$ and $\beta$ indicating a specific channel) or one channel per array on modules 3 and 4. The signals generated in coincidence by cosmic rays have been acquired using the oscilloscope without amplification and with a 40 GHz sampling rate, employing the constant fraction triggering technique as the time pick-off method. A typical signal amplitude is of several hundreds of mV for PMT and several tens of mV for SiPM, as can be seen in Fig.~\ref{fig:oscillo}, where the signals (as they appear on the oscilloscope display) are shown. 

\begin{figure}[ht!] \centering
\includegraphics[width=0.42\textwidth]{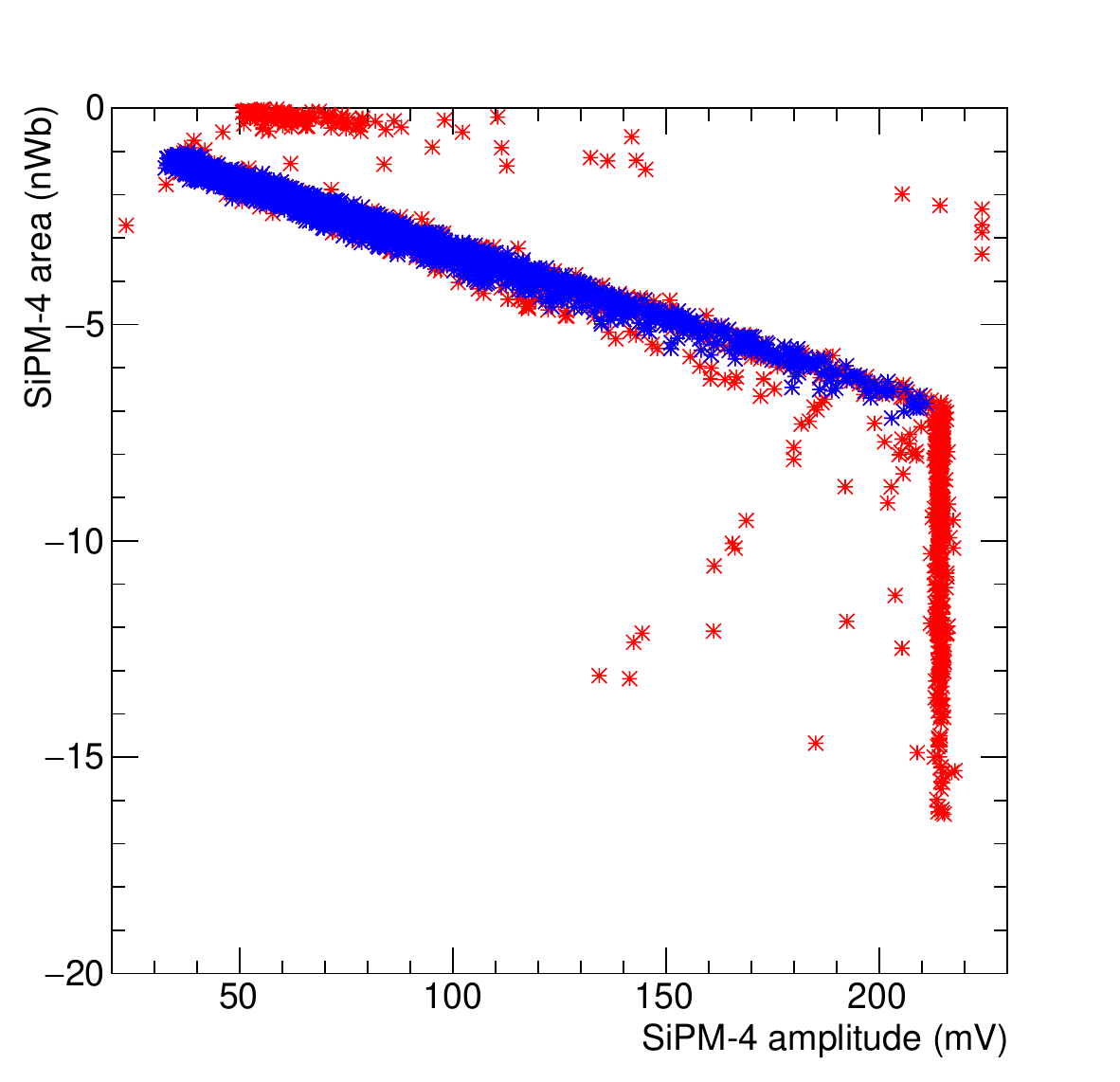}
\includegraphics[width=0.42\textwidth]{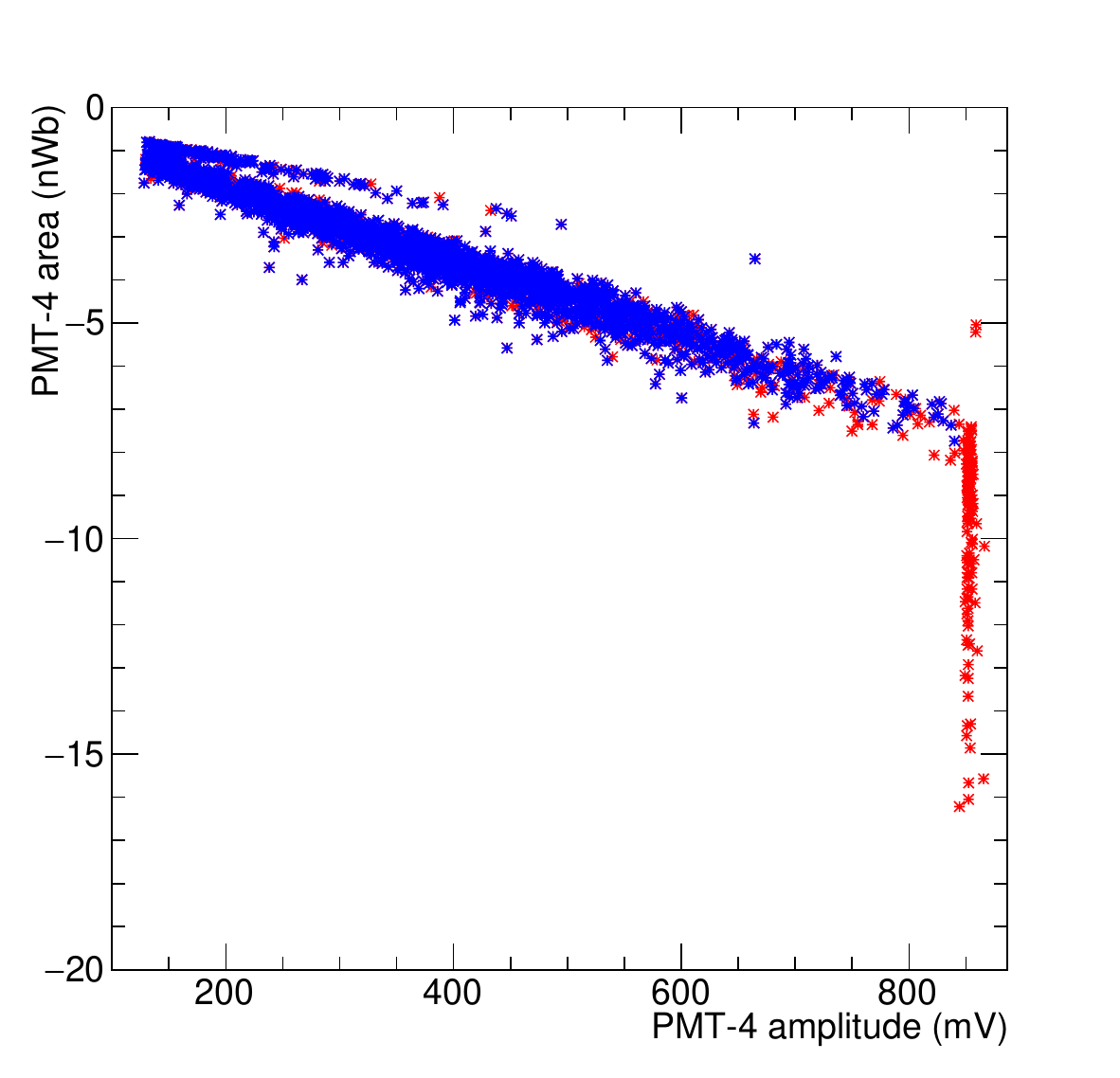}
\caption{Area versus amplitude of the signal for SiPM 4 (left) and PMT 4 (right).
The events with red markers are rejected (more detail in the text). Similar plots are available for SiPM 3 and PMT 3.} \label{fig:area_vs_amp}
\end{figure}

The PMT pulses are narrower ($\sim$20~ns) whereas the SiPM pulses are wider by a factor of four ($\sim$80~ns). Furthermore, it can be observed that the signal shape can be approximated with a triangle with the base indicating the duration. 
The area of the signals with respect to their amplitudes is shown in Fig.~\ref{fig:area_vs_amp} for SiPMs and PMTs. 

Both plots show a linearity of area versus amplitude meaning a quasi-constant duration, as expected. Thus the events of interest have been selected by fitting the linear region on these plots and by defining a selection range around the fit line. 
The accepted events are shown with blue markers, while the red ones are rejected. The latter are events with saturated signals on at least one channel. For SiPMs, also an electromagnetic noise~\footnote{The source of this noise was uniquely identified but it was not possible to eliminate.} was collected because of a lack of shielding of the custom electronic board used to pick up the signals. Such noisy signal was characterized by an area close to zero, has been removed off-line and did not affect the PMT signals.

\begin{figure}[ht!] \centering
\includegraphics[width=0.42\textwidth]{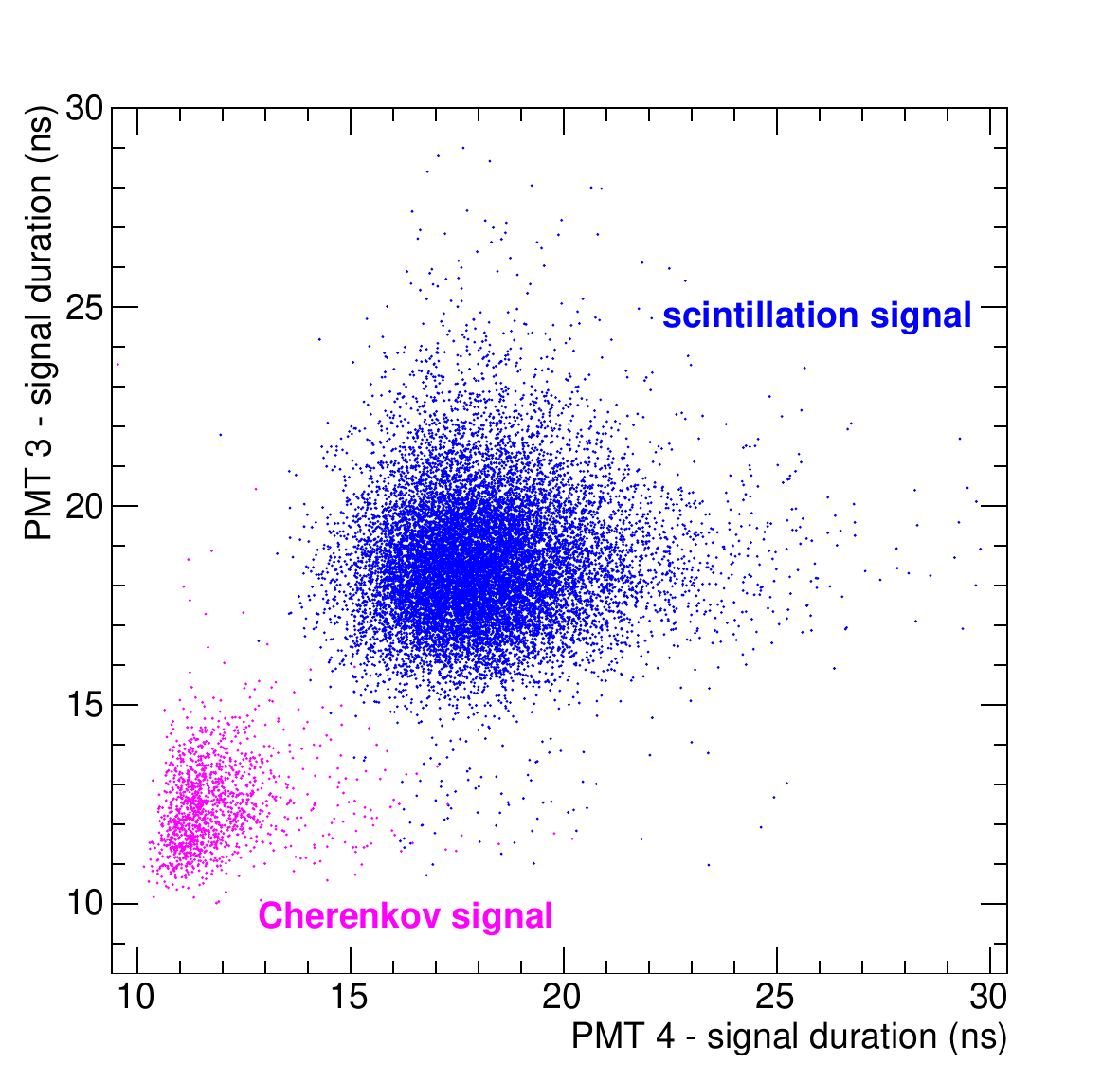}
\caption{Estimate of the signal duration for PMT~3 and PMT~4. The two separated clusters of events are attributed to photons generated by the Cherenkov (magenta) or scintillation (blue) effect.} 
\label{fig:chere_PMT}
\end{figure}

The PMTs exhibit the presence of two families of events, identified by bands with different slopes (Fig.~\ref{fig:area_vs_amp}, right). Considering the signal triangular shape, it results $area \sim (a/2) \times amplitude$, where the quantity $a$ is a rough estimate of the duration. It follows that the less slanted family of events is characterized by a faster development. The separation is more evident in Fig.~\ref{fig:chere_PMT}, where the $a$-quantities from both PMTs are shown together. The short signal duration is attributed to the Cherenkov effect occurring in the lucite light guides near the PMTs. The longer signal has a higher number of events, and matches the characteristics of the scintillation light produced by particles crossing the calorimeter. 

\begin{figure}[htbp]
\centering
\includegraphics[width=0.42\textwidth]{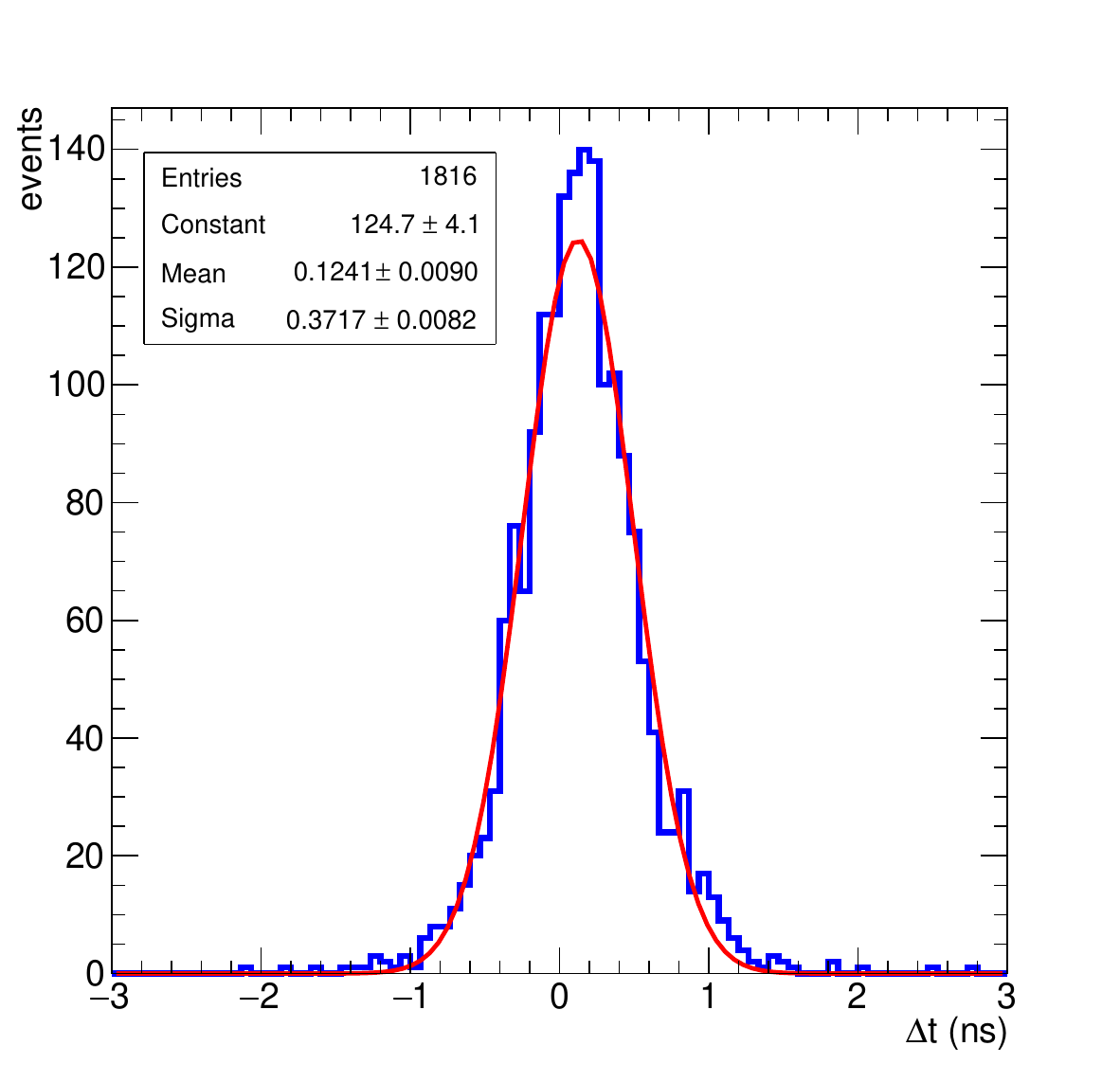}
\includegraphics[width=0.42\textwidth]{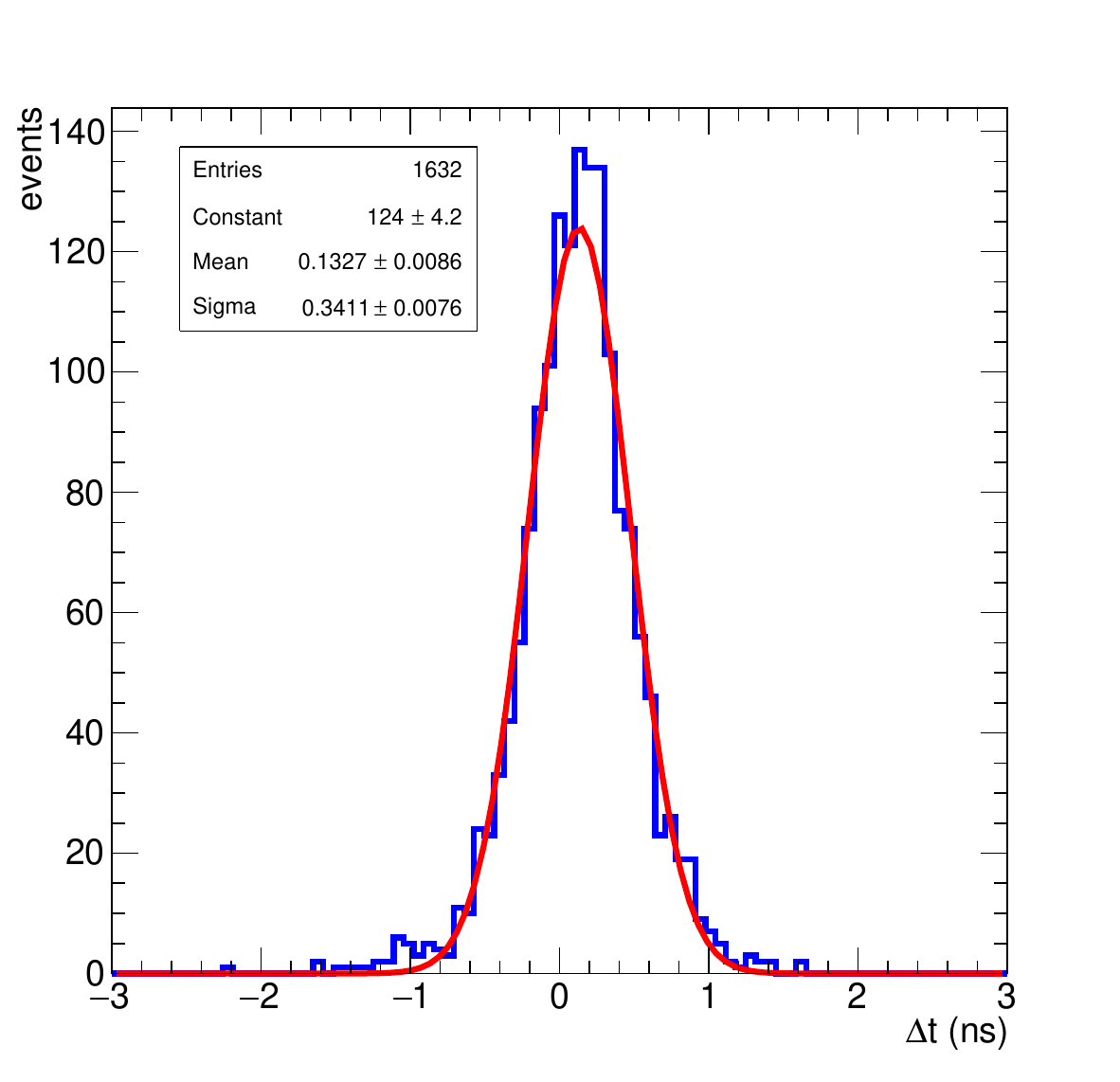}
\caption{Time difference for two channels of SiPM 3 (left) and SiPM 4 (right). The Gaussian fit (in red) is superimposed.}
\label{fig:time}
\end{figure}

\begin{figure}[ht!]
\centering
\includegraphics[width=0.42\textwidth]{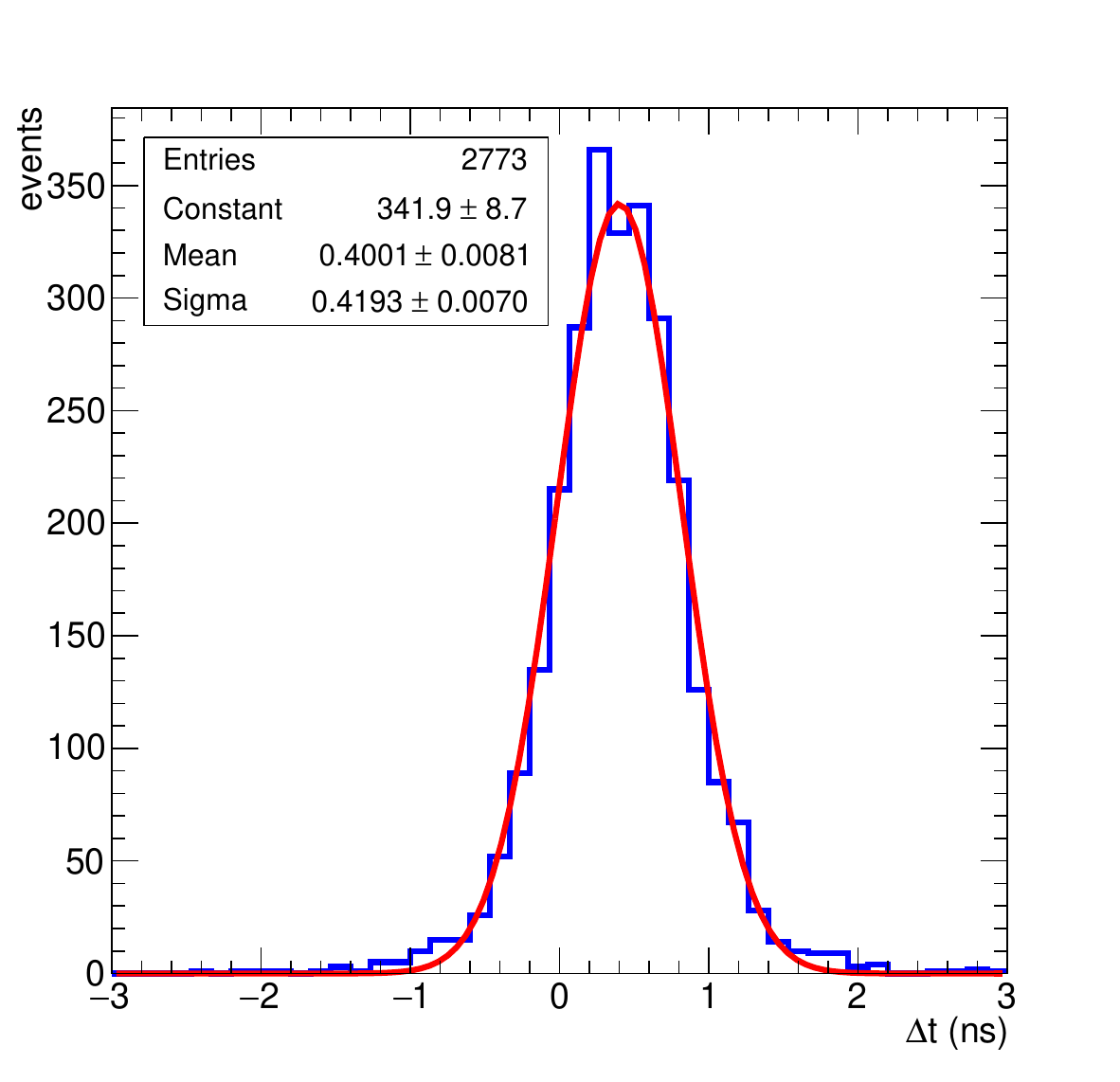}
\includegraphics[width=0.42\textwidth]{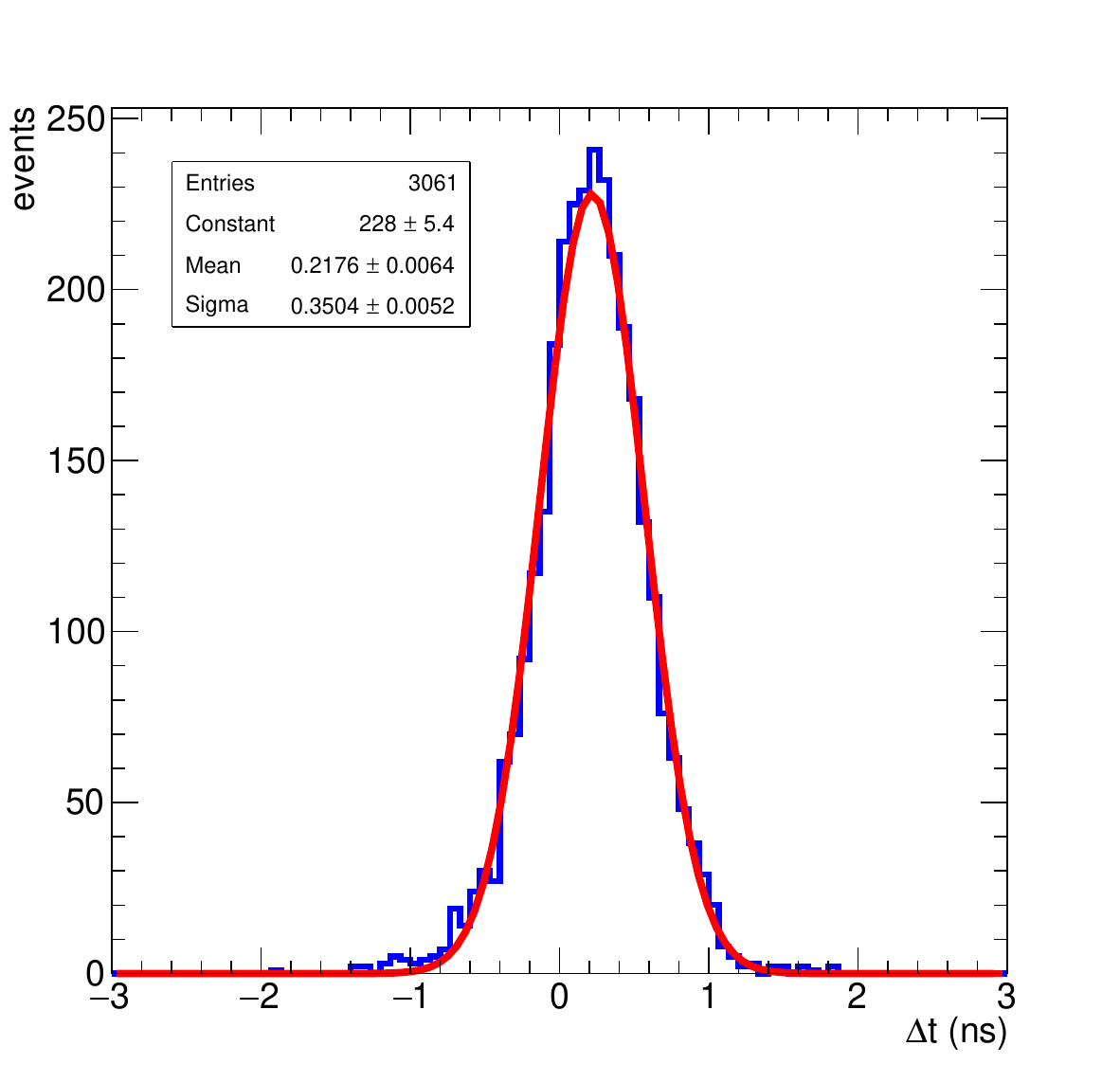}
\caption{Left: time difference for SiPM 3 and SiPM 4. Right: time difference for 
PMT 3 and PMT 4. The distributions are fitted with a Gaussian function (in red).}
\label{fig:time_PMT}
\end{figure}

The time difference at 50\% of the signals is shown in Fig.~\ref{fig:time} for two channels of the same SiPM matrix, and in Fig.~\ref{fig:time_PMT} for SiPMs (left) and PMTs (right) placed on different ECAL modules (3 and 4), after the event selection described above. For PMTs, only the scintillation light is taken into account. 


For two channels of the same SiPM array (Fig.~\ref{fig:time}), the distribution mean value is due to an offset ($\sim 128$ ps), introduced by the custom electronic board used to pick up the signals. Since each SiPM in the array gives an independent signal readout, the intrinsic timing resolution ($\sigma_t$) of the single SiPM channel is $\sigma_t = \sigma_{\Delta t}/\sqrt2$.

In the case of detectors coupled with different ECAL modules (Fig.~\ref{fig:time_PMT}), the mean value depends on the average time of flight of a relativistic particle from module 3 to module 4 and, for the SiPMs, also on the electronics offset and the delay introduced by the adapter on SiPM~4. Consequently, the $\sigma_{\Delta t}$ of these distributions is affected by the additional jitter due to the various particle paths (geometrical jitter). The intrinsic timing resolution is estimated by simulating the experimental setup and a muon flux proportional to cos$^2\theta$, with $\theta$ being the zenith angle~\cite{bib:cos2}. The estimation consists of looking for the intrinsic timing resolution which reproduces the measured $\sigma_{\Delta t}$ when combined with the geometrical jitter. All the measurements of the timing resolution are summarized in Table~\ref{tab:det}. It can be concluded that the PMTs exhibit a slightly better timing resolution than the SiPM.

\begin{table}[htbp]
\centering
\caption{Timing resolution for SiPMs and PMTs. The $\sigma_{\Delta t}$ and its error are derived from the fit of distributions in Figs.~\ref{fig:time} and \ref{fig:time_PMT}. The $\sigma_t$ is the intrinsic timing resolution of the single detector.}\label{tab:det} 
\smallskip
\begin{tabular}{l c c} 
\hline
                          & $\sigma_{\Delta t}$ (ps) & $\sigma_t$ (ps) \\ \hline
SiPM 3 $\alpha$ - $\beta$ & 372$\pm$8          & 263$\pm$6 \\
SiPM 4 $\alpha$ - $\beta$ & 341$\pm$8          & 241$\pm$6 \\
SiPM 3$\alpha$ - SiPM 4$\alpha$          & 419$\pm$7          & 269$\pm$5 \\
PMT 3 - PMT 4             & 350$\pm$5          & 217$\pm$4 \\ \hline  
\end{tabular} 
\end{table}

The same measurement procedure was repeated at four different temperatures in the range from $\sim17^\circ$C to $\sim27$°C, measured by a sensor placed near the SiPM matrix. It has been verified that the SiPM timing resolution is not significantly affected by the temperature variation in this range.

\begin{figure}[htbp]
\centering
\includegraphics[width=0.42\textwidth]{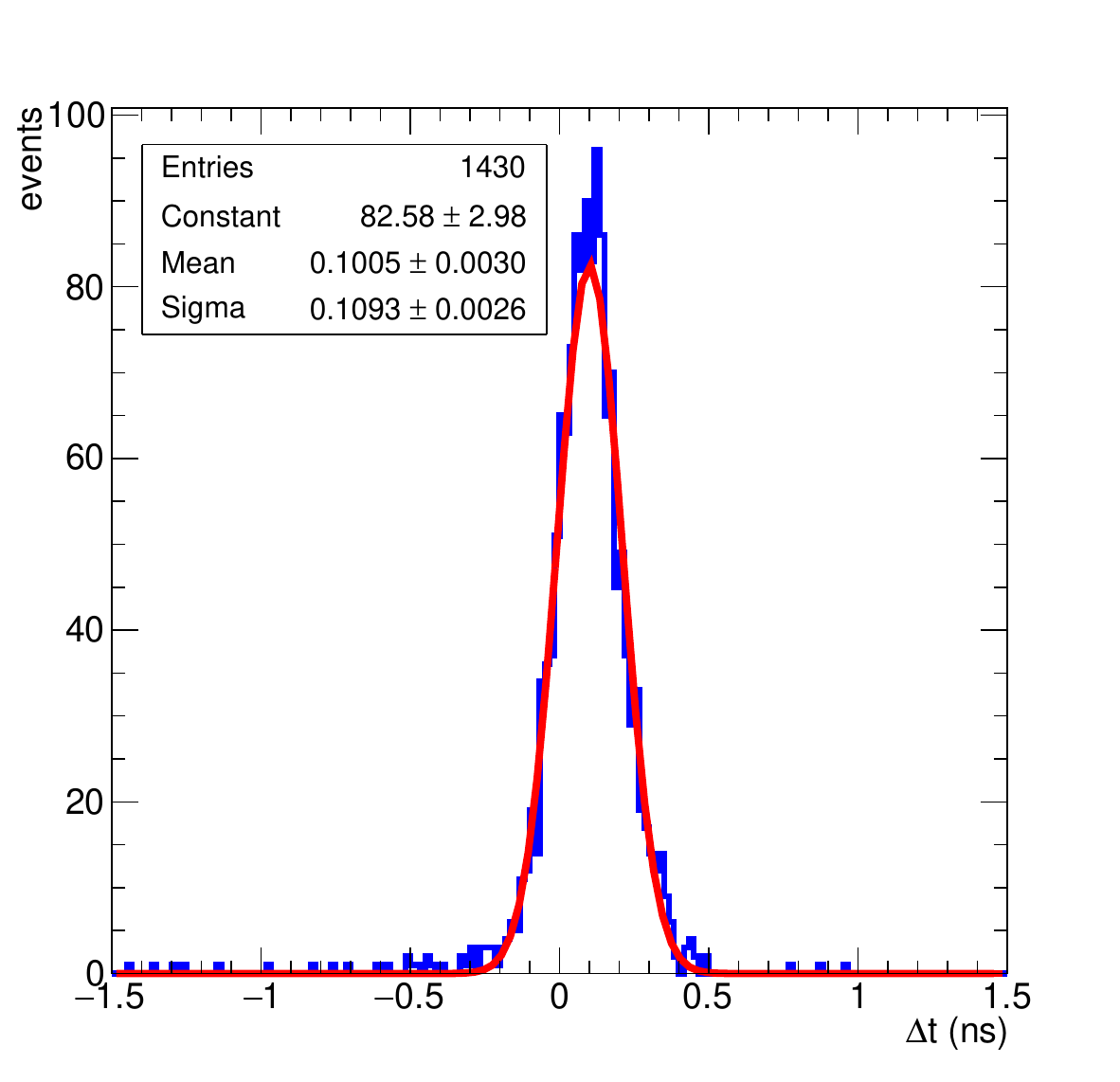}
\caption{Time difference for PMT~3 and PMT~4, considering only Cherenkov light.}
\label{fig:hchere}
\end{figure}

The $\Delta t$ distribution for Cherenkov events detected in the PMTs is shown in Fig.~\ref{fig:hchere}, fitted with a Gaussian function. The mean value and the $\sigma_{\Delta t}$ are smaller than what is obtained when the scintillation light is considered due to intrinsic characteristics of the scintillation process. 
Specifically, in this case, $\sigma_{\Delta t}$ = 109 $\pm$ 3~ps, which results in a timing resolution $\sigma_t \sim$ 70~ps. 
The Cherenkov effect was not observed in the SiPMs collected data. This is probably related to the fact that the SiPM signal is not amplified. 

The timing resolution of the KLOE calorimeter with PMT read-out (used in this work) depends on the deposited energy ($E$), according to the following formula~\cite{bib:kloe}:

\begin{equation}
\sigma_{t} \sim 54~ps / \sqrt{E~(\rm{GeV})}. 
\label{eq:resKLOE}
\end{equation}

\begin{figure}[htbp]
\centering
\includegraphics[width=0.42\textwidth]{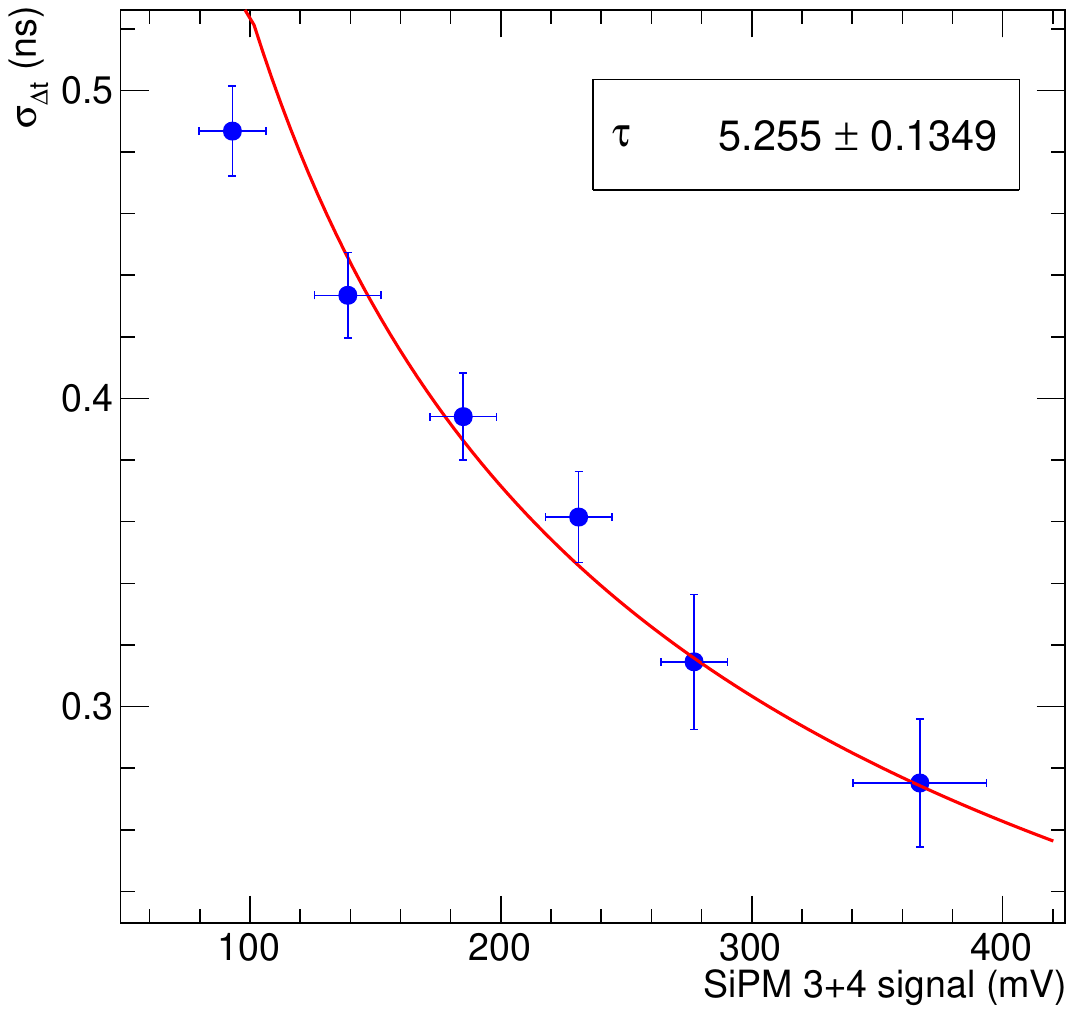}
\includegraphics[width=0.42\textwidth]{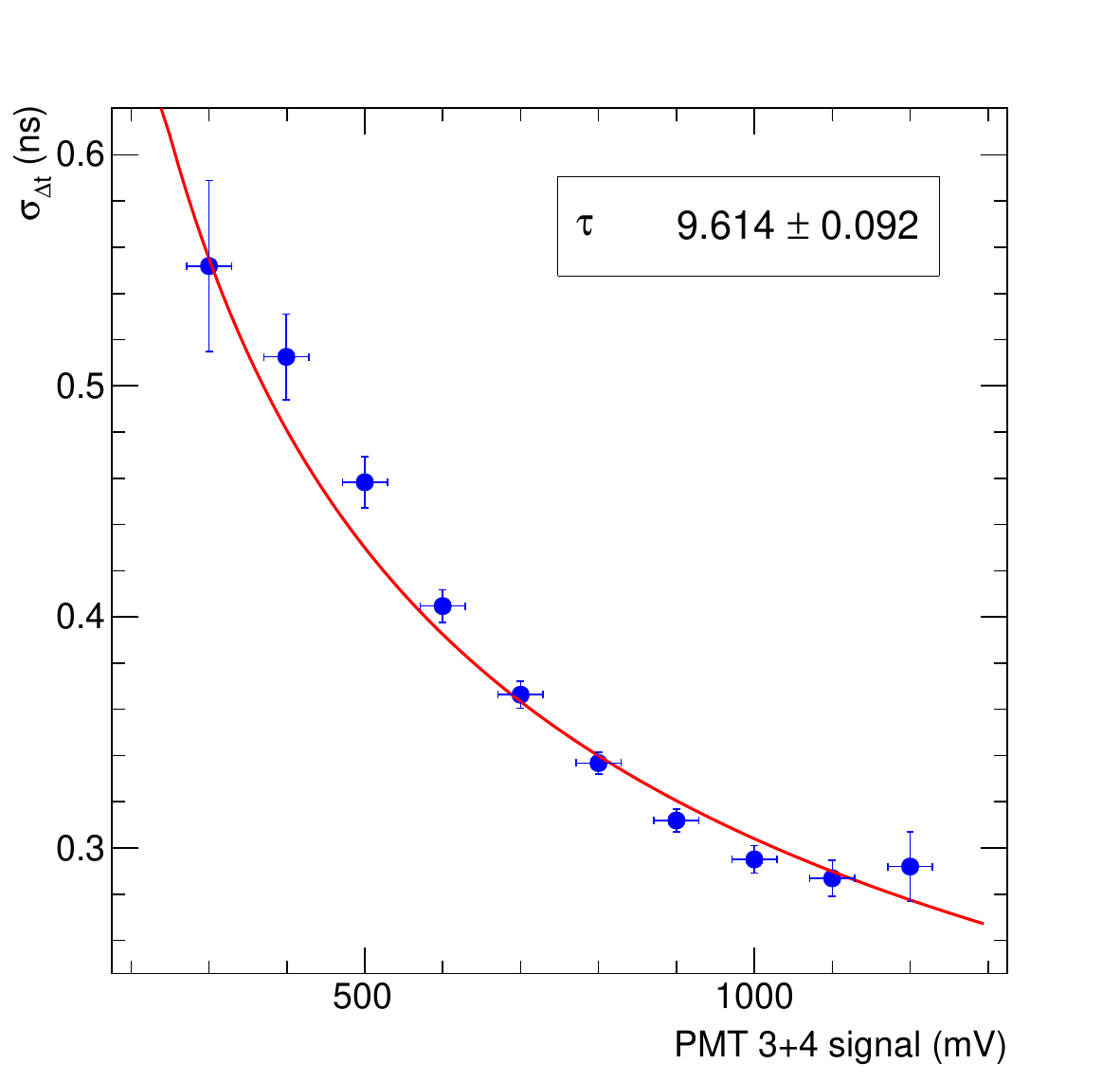}
\caption{Timing resolution as a function of the signal for SiPMs (left) and PMTs (right).
The red curve represents the fit function reported in the text (see formula~\ref{eq:ff2002}).}
\label{fig:formula2002}
\end{figure}

\noindent A similar behavior has been observed in this study for SiPMs and PMTs, when the signal amplitude is used instead, being it related to the deposited energy. The results are shown in Fig.~\ref{fig:formula2002}, and are well described by the formula:

\begin{equation}
\sigma_{\Delta t} = \tau / \sqrt{signal~(\rm{mV})},
\label{eq:ff2002}
\end{equation}

\noindent where $\tau$ is the fit parameter. Although the value of $\tau$ cannot be compared with 54 ps because signal (in mV) and energy (in GeV) are different quantities, they are typically linearly related and the similarity of equations~(\ref{eq:resKLOE}) and (\ref{eq:ff2002}) confirms the dependence of timing resolution on energy (and signal amplitude).

In conclusion, the timing resolution was computed and compared for PMTs and SiPMs. When the scintillation light is considered, the two photodetectors show similar results, with the PMT performing slightly better. Additionally, PMTs are also sensitive to Cherenkov light, exhibiting in this case a better timing resolution than the scintillation one.

\subsection{Consistency check on the measurement method}\label{sec:check}
\begin{figure}[htbp] \centering 
\includegraphics[width=0.45\textwidth]{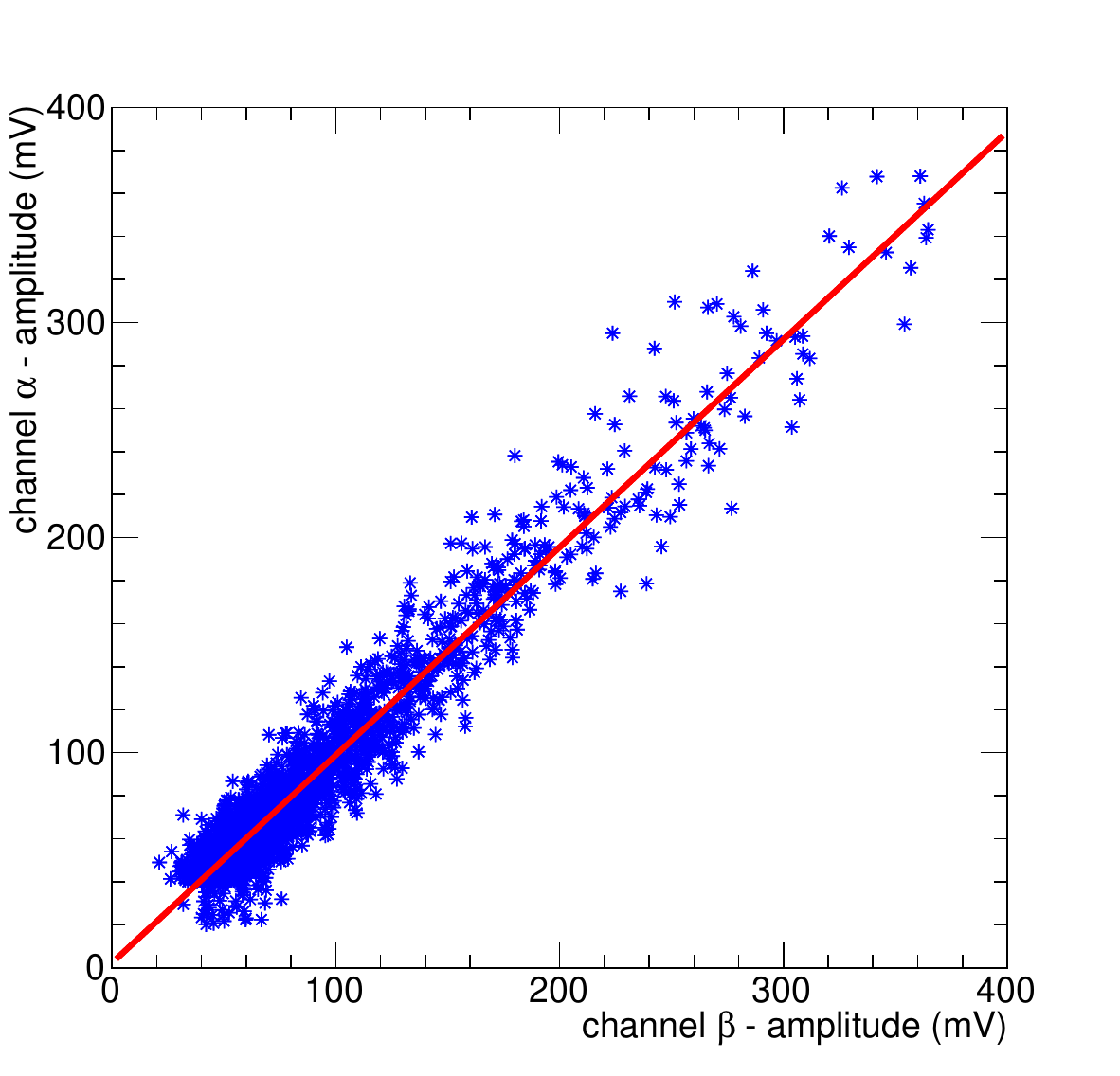}
\caption{Signal amplitude of channel $\alpha$ versus amplitude of channel $\beta$ of the same SiPM array.}
\label{fig:amp_vs_amp} \end{figure}

Some checks were made during the process, in order to validate the performed measurements.
The first consistency check is shown in Fig.~\ref{fig:amp_vs_amp}, where the amplitude of signals taken from two separate channels of the same SiPM matrix are compared. As expected, the distribution of events is well described by a bisector. This result confirms the uniformity of the photon collection event-by-event on the SiPM array. This uniformity was already observed on a statistical basis (see Fig.~\ref{fig:uniform}, right).

\begin{figure}[htbp] \centering
\includegraphics[width=0.45\textwidth]{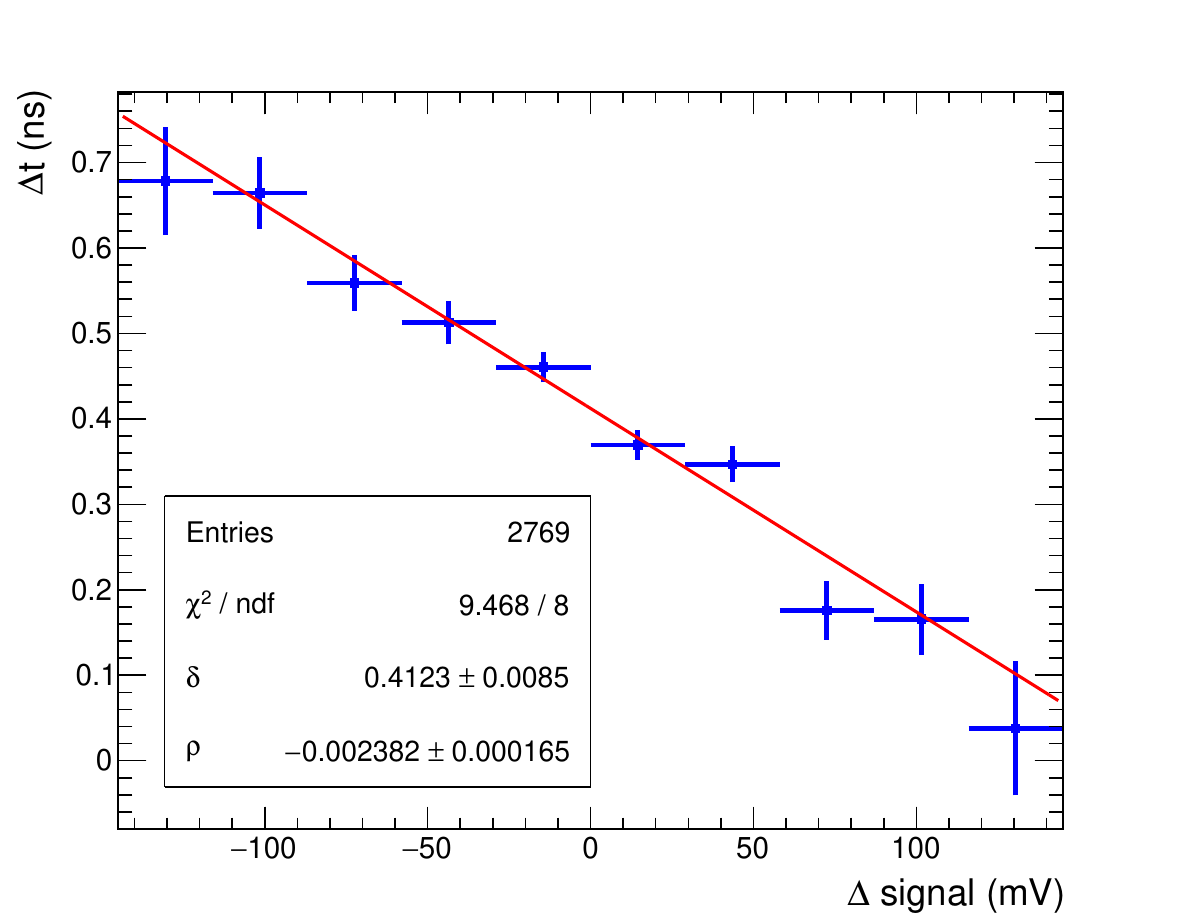}
\includegraphics[width=0.45\textwidth]{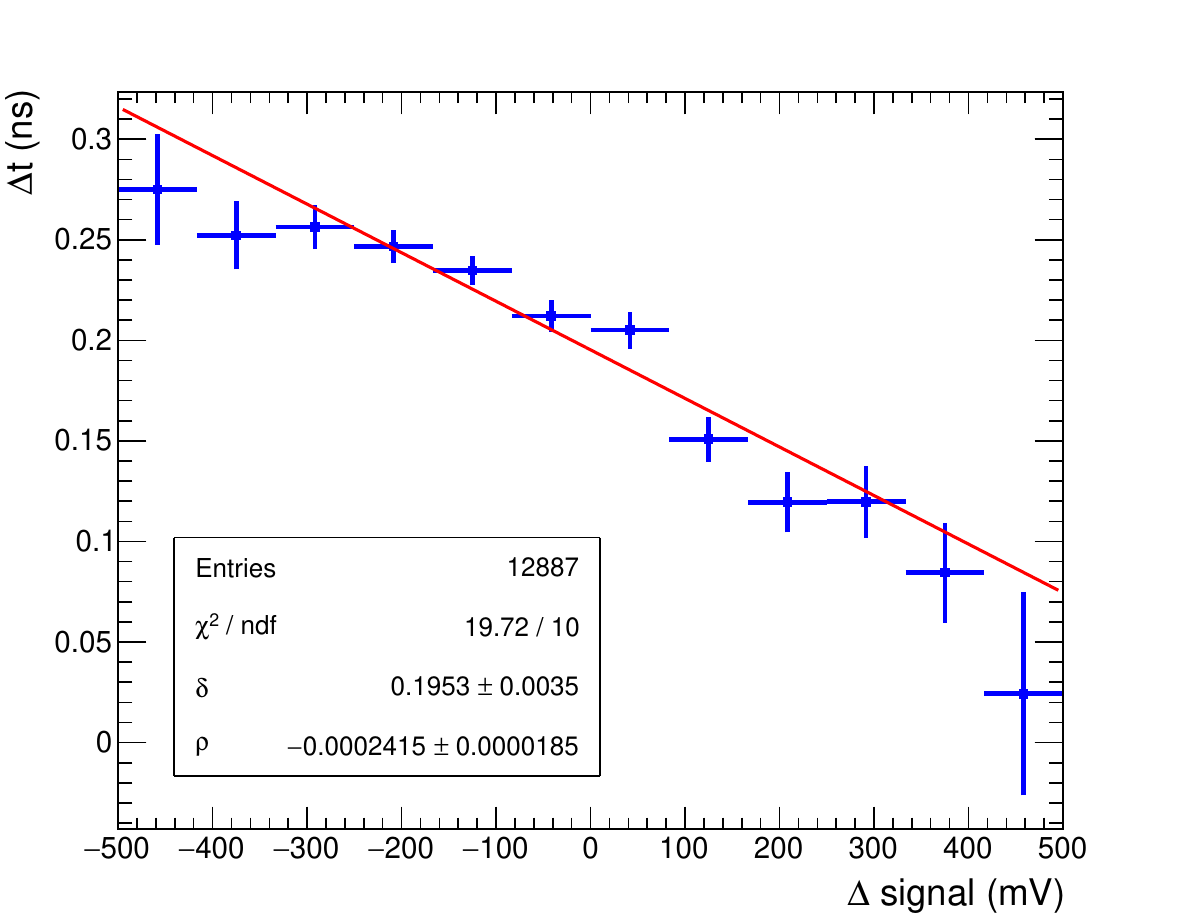}
\caption{Time difference vs signal amplitude difference (profile). Left: SiPM. Right: PMT.}
\label{fig:diff_vs_diff} \end{figure}

Another check can be performed by looking at the difference in time ($t_4 - t_3$) and signal amplitude ($signal_4 - signal_3$) on different ECAL modules. 
These quantities are related because both increase with the track slope.
The complete description of this dependence must take into account the exponential reduction of the number of detected photons (and the collected signal) with the path in the calorimeter bar, depending on the attenuation length ($\sim 325$~cm for Pol.Hi.Tech-0046 fibers~\cite{bib:fiber}). Since the light path is on average much smaller than the attenuation length, it is possible to approximate this exponential dependence with a linear one. This is confirmed from the profiles in Fig.~\ref{fig:diff_vs_diff} where a linear relation between $\langle \Delta signal \rangle$ and $\langle \Delta t \rangle$ is visible for SiPMs and PMTs.

\section{Conclusions}\label{sec:conclu}
A cut-out of the KLOE calorimeter has been equipped with SiPMs and PMTs in order to compare their performance for a potential application in an element (SAND) of the Near Detector complex of the future DUNE experiment. The PMT coupling was optimized in the KLOE calorimeter 
and it was kept unchanged for this study. Various optical couplings have been tested for adapting the SiPMs with the KLOE light
guides and it was verified that the partial coverage of the light guide does not prevent reaching a proper readout of the signal.
Efficiency and timing resolution have been measured and compared for both detectors. It was found that the SiPM noise cannot be
neglected in the efficiency estimate. Specifically, the noise rejection implies a non-negligible reduction of the possible SiPM
efficiency making it lower by a few percent with respect to that of the PMTs. For the timing resolution, it was observed that PMTs 
reach a value lower than 220~ps for the scintillation signal, while SiPMs exhibit a resolution above 240~ps. Even if the difference is small, PMTs seem to perform better. 

In conclusion, the difficulties in coupling SiPMs to ECAL without major mechanical changes, the lack of improvement, the cost, and the necessary commissioning time, advise against the substitution of the 4880 available and tested PMTs with new SiPMs in the KLOE calorimeter to be reused in the Near Detector complex in DUNE. Nonetheless, the results from this study do not exclude the possible use of SiPMs for other calorimetry applications.

\acknowledgments
We would like to thank Stefano Miscetti (INFN, Laboratori Nazionali di Frascati) who delivered us the stack of lead foils and scintillating fibers. We thank also 
Antonio Di Domenico (Sapienza, Rome) and Alessandro Montanari (INFN, Sezione di Bologna) for their fruitful comments and suggestions.





\end{document}